\documentclass[%
  reprint,            % use "preprint" for single-column draft
  aps,                % American Physical Society
  prd,                % target journal: Physical Review D
  superscriptaddress, % common for multi-institution papers
  nofootinbib         % optional: move footnotes into text
]{revtex4-2}

\usepackage{graphicx}% Include figure files
\usepackage{dcolumn}% Align table columns on decimal point
\usepackage{bm}% bold math
\usepackage{lipsum}
\usepackage{amsmath}
\usepackage{xcolor}

\begin{document}

\preprint{APS/123-QED}

\title{Finding the Edge of Chaos in a Ferromagnet: \\ Quantifying the ``Complexity'' of 2D Ising Phase Transitions with Image Compression}% Force line breaks with \\

\author{Cooper Jacobus}
\email{jacobus@stanford.edu}
\affiliation{Department of Physics, Stanford University}%
%\affiliation{Kavli Institute for Particle Astrophysics and Cosmology, Stanford, CA}%

\date{\today}% It is always \today, today,

\begin{abstract}
The data-driven characterization of the ``complexity'' present in dynamical systems remains an open problem with broad applications across the physical sciences. We investigate the ``structural complexity'' of the 2D ferromagnetic Ising model, a paradigmatic system exhibiting a second-order phase transition at a certain critical temperature which is often cited as a canonical example of complex morphology. We define a quantitative metric for this structural complexity, $\mathcal C_s$, through the lens of algorithmic information theory by approximating the Kolmogorov complexity of lattice configurations via standard lossless image compression algorithms. We regularize our proposed metric, $\mathcal C_s$, by comparing the compressibility of a configuration to that of its pixel-wise sorted and randomly shuffled counterparts. We arrive at a definition of $\mathcal C_s$ as a product of two components representing the systems departure from perfect order and disorder respectively which we then plot as a function of temperature.  Our numerical simulations reveal a distinct peak in $\mathcal C_s$ at the known critical temperature $T_c$. This result demonstrates that such information-theoretic measures can act as sensitive, model-agnostic indicators of criticality, directly quantifying the emergence of complex structure at the boundary between order and chaos, opening the door to data-driven applications in domains where analytic solutions are unavailable.
\end{abstract}

\maketitle

\section{Introduction}
\label{sec:introduction}

The emergence of complex, large-scale structures from simple, local rules is a unifying theme across the natural sciences. Many physical systems exhibit the seemingly bizarre phenomenon in which intricate, adaptive, and often unpredictable macroscopic behavior arises from simple, local interactions among a system's components \cite{Mitchell2009}. Such systems, spanning from gene regulatory networks and neural dynamics to social structures and ecosystems, characteristically operate in a so-called ``critical'' regime between featureless order and randomness \cite{Langton1990}. At this ``edge of chaos," systems are thought to possess a maximal capacity for information processing, adaptation, and computation. Understanding the principles that govern this behavior and the conditions under which they apply remains an unresolved scientific challenge. 

This intuitive notion of ``complexity'' has inspired a rich history of theoretical work, including many pioneering developments in the past half-century. 
%These efforts have sought to move beyond traditional macroscopically-defined quantities like thermodynamic entropy, and toward the identification of macrostate-localized structure. 
These efforts have sought to complement macroscopic quantities like entropy by providing insight into a system's internal structure or history. 
The term ``complexity'' is used in many domain-specific contexts to mean a variety of different things. Several more formalized definitions include: ``effective complexity," which measures the descriptive length of a system's set of regularities \cite{GellMann1996}; ``thermodynamic depth," which measures the amount of computation required to build a state from a simple origin \cite{Lloyd1988}; and the ``statistical complexity" of computational mechanics, which quantifies the minimum information required about a system's past to predict its future \cite{Crutchfield1989}.

While conceptually powerful, these formal metrics often remain difficult or computationally intractable to apply directly to high-dimensional data without significant model-specific adaptation.
However, the development of massive scientific datasets and unprecedented computational power may enable to application of data-driven proxies for the complexity metrics. From exabyte-scale astrophysical surveys and genomic catalogs to neurological imaging and global climate modeling, modern science is increasingly defined by the ability to gather vast quantities of information \cite{Hey2009}. A central challenge of this new paradigm is not merely the storage of this data, but its interpretation. Hidden within these massive datasets are the signatures of complex, emergent, often unpredictable patterns. The automated discovery and characterization of these emergent structures is an unexplored frontier, demanding new, scalable methods of analysis that can operate without model-specific assumptions.

The development of new, practical, model-agnostic, and broadly applicable complexity metrics for image-structured data holds significant promise across a variety of unique disciplines. It could, in Material Science for instance, be used to quantify the micro-structural complexity and behavior of novel materials, beyond what can be deduced from obvious crystalline symmetries (e.g. \cite{DeCost2017, Martiniani2019}). In Medicine, it could be employed to automatically quantify tissue dysplasia and heterogeneity in medical imaging data (e.g. \cite{Irshad2014}) and provide a meaningful predictor of outcomes like tumor malignancy \cite{Gatenby}. In domains like Neuroscience and Astrophysics, wherein previously unimaginable volumes of data are being collected, and where frontier problems are increasingly defined by their complex behavior, computational tools for measuring complexity could prove useful for anything from cartography of the Interstellar Medium (e.g. \cite{Elmegreen2004}) to the quantitative description of neural activation maps (e.g. \cite{Friston1994}).

\begin{figure*}
  \centering
  \includegraphics[width=\textwidth]{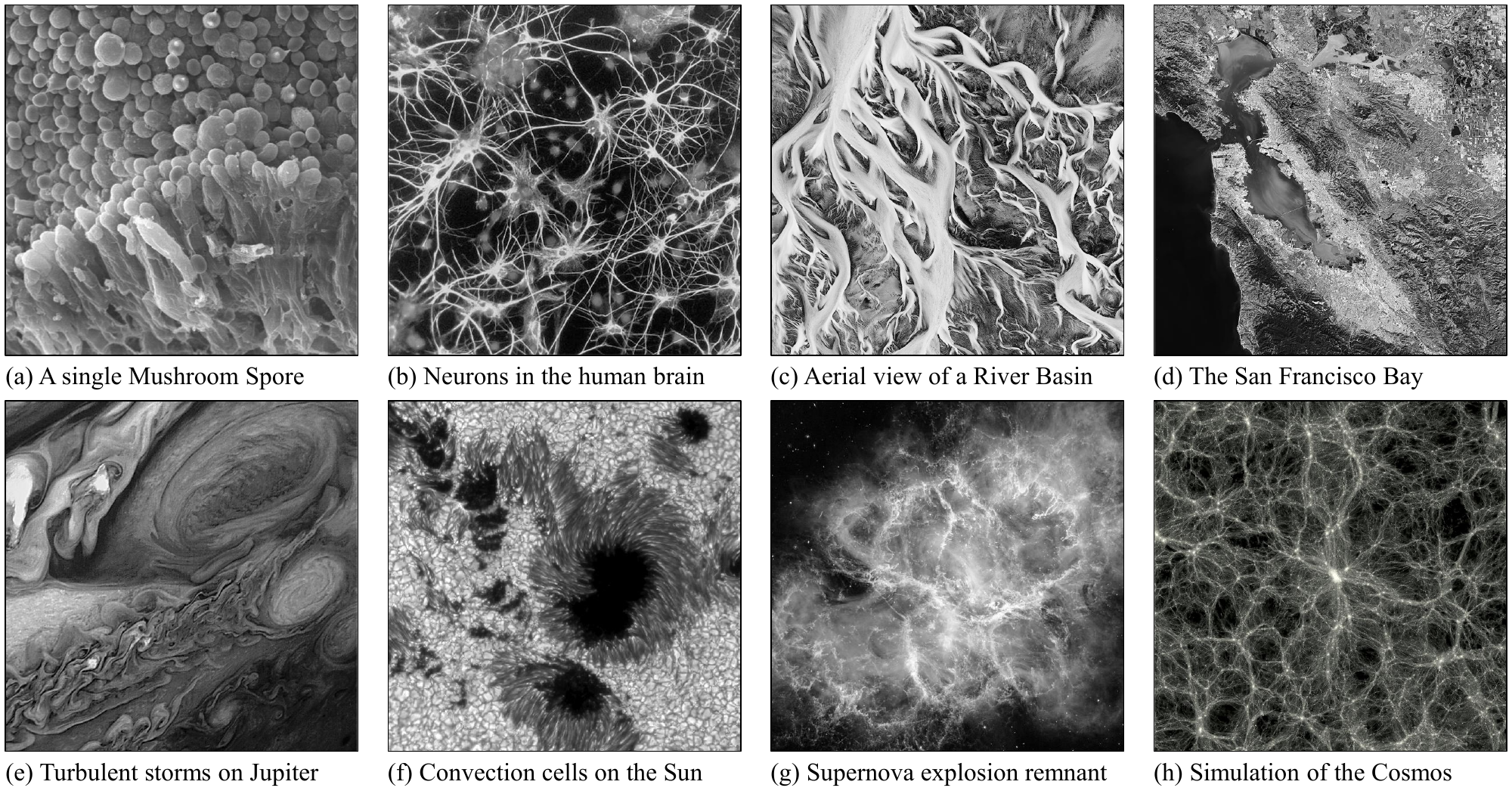}
  \caption{Example images of ``structurally complex'' systems from various domains over many orders of magnitude in scale. \\ Image Credits: (a) Dartmouth College, (b) Nancy Kedersha, (c) Emmanuel Coupé, (d), ESA, (e,f,g) NASA, (h) Volker Springel.}

  \label{fig:complex_structures}
\end{figure*}

\newpage

Towards these ends, we would like to explore a measure that is both computationally tractable, system-agnostic, and intuitively meaningful, and that could serve as a more broadly applicable quantification for complexity. 
To define such a metric, in this work, we argue that a practical and universal approach can be found in algorithmic information theory, which describes the complexity of an object with the length of its shortest possible description, or Kolmogorov complexity \cite{Kolmogorov1965}. We demonstrate that a practical, compression-based approximation of this measure, when properly regularized, can reliably extract meaningful insight and serve as a detector of criticality in image-structured data.

While formally uncomputable \cite{Turing1936}, the Kolmogorov complexity of any signal can be practically approximated using modern lossless compression algorithms \cite{Ziv1977}. This compression-based methodology offers a model-agnostic, computationally feasible proxy for descriptive complexity, allowing us to directly quantify the presence of any structure within data. To leverage the descriptive, pattern-recognizing power of image compression, we introduce a normalized ``structural complexity'' metric, $\mathcal C_s$, which isolates the compressibility arising from non-trivial spatial correlations from that due to simple compositional bias. We define $\mathcal C_s$ as the geometric mean of two components, $\mathcal O_s$ and $\mathcal D_s$, representing the distance between the image's compressed size and a disordered and ordered baseline, respectively. We motivate that this parameter $\mathcal C_s$ extracts an intuitive measure of the structural content of a given image.

%A physical system which undergoes a phase transition at some critical point may, for example, display very ordered behavior below that point and very disordered behavior above it, while producing complex self-similar behavior around the critical point itself. While established and well-defined thermodynamic quantities like entropy effectively measure disorder, they fail to capture the intuitive notion that a system is most ``complex'', exhibiting a rich hierarchy of structures, when poised somewhere between absolute order and randomness \cite{Langton1990}. This intermediate regime demands a more nuanced quantitative framework.

\newpage
To rigorously test our proposed framework, we apply it to a paradigmatic model of ``complex'' behavior: the two-dimensional (2D) Ising model. In physics, some of the most quintessential manifestations of this ``complex'' phenomenology occur around critical phase transitions, which the Ising model replicates, where microscopic interactions give rise to intricate macroscopic behavior and often produce a rich, hierarchical structure. 
By treating simulated lattice configurations as digital images, we measure their compressibility as a function of temperature. Since we already understand its different behavioral regimes, we can use the 2D Ising model as a canonical benchmark for the framework we develop. We expect that a meaningful structural complexity metric $C_s$ should be maximized precisely at the model's known critical temperature, where the system's fractal-like, scale-invariant structure is most pronounced and long-range interactions can manifest. By analyzing the compressibility of many simulated spatial configurations over a range of $T$, we show that our metric exhibits a sharp, unambiguous peak around the critical temperature, giving credence to the broader applicability of our approach.

This paper is structured as follows: Section~\ref{sec:model} reviews the 2D Ising model and our Monte Carlo simulation methods. Section~\ref{sec:quantifying_complexity} details the theoretical basis and practical implementation of our compression-based complexity metrics. Section~\ref{sec:results} presents and discusses our numerical results. Finally, Section~\ref{sec:discussion} discusses the broad implications of these findings and the potential for applying this methodology to other domains.

\begin{figure*}
  \centering
  \includegraphics[width=\textwidth]{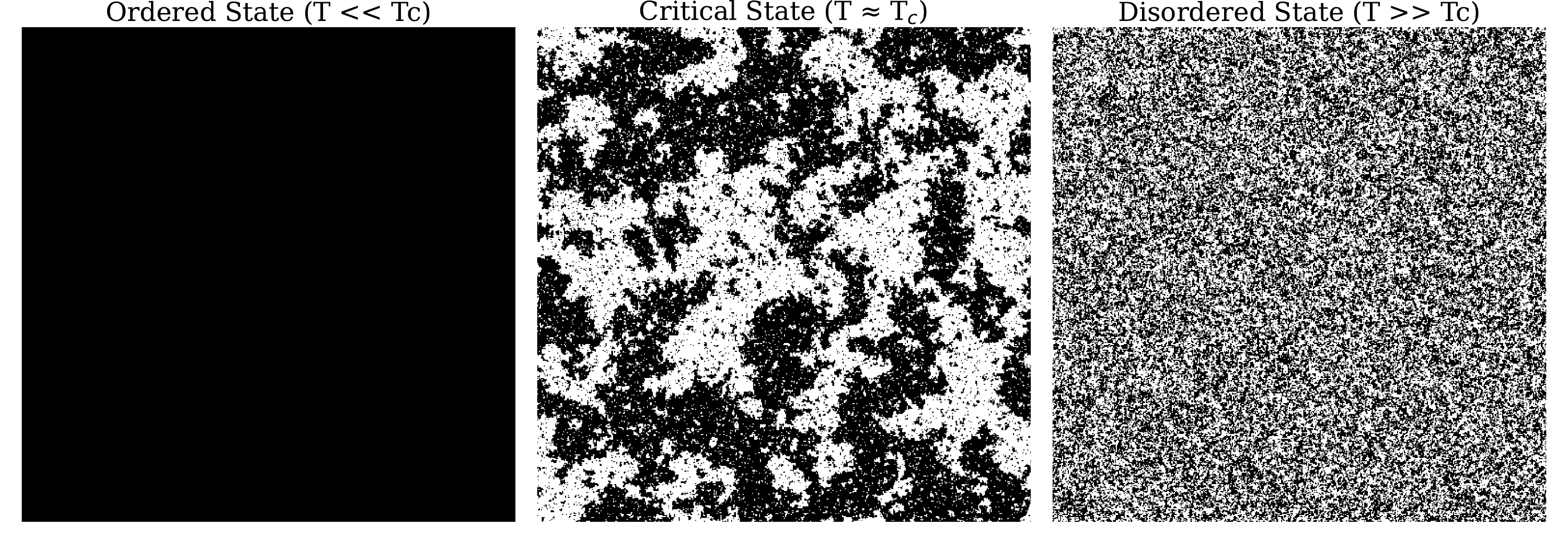}
  \caption{Visualization of thermalized 2D Ising lattices generated by our simulation method from initially random conditions with width $N=256$ cells. Samples illustrate the ordered, critical, and disordered regimes relative to the critical temperature. }
  \label{fig:Ising_images}
\end{figure*}

\newpage
\section{The 2D Ising Model}
\label{sec:model}

In the study of complex systems, and in physics more generally, it is often useful to study simplified ``toy models'' which isolate the behavior of interest while otherwise abstracting away the irrelevant details. The Ising model stands as perhaps the most canonically influential such ``toy model'' and will serve as an established benchmark to validate the methodology and definition proposed in this paper. To make this work more accessible to non-physicists, we review the definition and behavior of this model here and discuss our simulation methodology.

\subsection{Model Definition}

The Ising model, originally conceived to describe ferromagnetism, is a foundational model in statistical physics for studying phase transitions and emergent behavior \cite{Ising1925}. In its two-dimensional form on a square lattice, the Ising system consists of a grid of $N \times N$ sites, with each site $i$ occupied by a spin $s_i$ that can point either up ($+1$) or down ($-1$).  The system's behavior is driven by an energetic preference for neighboring spins to be aligned with one another. The total energy of a given spin configuration, $\{s_i\}$, is described by the Hamiltonian:
\begin{equation}
    H(\{s_i\}) = -J \sum_{\langle i,j \rangle} s_i s_j,
    \label{eq:hamiltonian}
\end{equation}
where the sum runs over all unique pairs of nearest-neighbor spins $\langle i,j \rangle$. The parameter $J$ is the coupling constant; for a ferromagnetic system ($J>0$), parallel spins ($s_i = s_j$) contribute $-J$ to the energy, while anti-parallel spins contribute $+J$. Thus, the local interaction rule energetically favors alignment. 
Hereafter, all energy and temperature scales are expressed in units where $J=1$ and the Boltzmann constant $k_B=1$.

\subsection{Behavior \& Thermodynamic Regimes}

The macroscopic behavior of the Ising model is a direct consequence of the competition between two fundamental thermodynamic forces: the tendency to minimize energy (favoring order) and the tendency to maximize entropy (favoring disorder). The temperature, $T$, serves as the control parameter that mediates this competition. The most interesting behavior is exhibited around a known intermediate ``critical'' temperature, $T_c$, where the energetic and entropic forces are perfectly balanced. This critical temperature thus divides the system's behavior into three qualitatively distinct regimes, as illustrated in Fig.~\ref{fig:Ising_images} above. These are described as:

\begin{enumerate}
    \item \textbf{Low Temperature ($T \ll T_c$):} The drive to minimize energy dominates. Thermal fluctuations are insufficient to flip spins against the ordering preference of their neighbors. The emergent state is simple and predictable. The system settles into a globally ordered phase where nearly all spins align. 
    \item \textbf{High Temperature ($T \gg T_c$):} Thermal energy dominates. The energetic cost of a misaligned spin is negligible compared to the entropic gain of randomness. Entropy is maximized as spins orient randomly, resulting in an absolutely disordered state.
    \item \textbf{Critical Temperature ($T \approx T_c$):} The system exists at a delicate balance between energy and entropy. and therefore cannot commit to either global order or complete disorder. The result is a state of continuous scale-invariant fluctuation, which leads to the formation of correlated spin clusters of similar morphology on all possible length scales.
\end{enumerate}
For the 2D square lattice, the exact critical temperature was famously solved by Onsager: $T_c = 2J / [k_B \ln(1 + \sqrt{2})]$ or about $T_c \approx 2.269 \space J/k_B$ in dimensionless units \cite{Onsager1944} .

\newpage
\subsection{Simulation Procedure}

To generate representative spin configurations from the Boltzmann distribution, $P(S) \propto \exp(-H(\{s_i\})/k_B T)$, we employ a traditional Monte Carlo method, updating the spin values according to a Markov process which is sensitive to the systems change in energy. The Markov chain steps can be generated by a variety of algorithms. Most intuitively, the Metropolis algorithm implements simple, local update scheme wherein a flip is proposed for a single randomly chosen spin and accepted or rejected according to the energetic cost \cite{Metropolis1953}. A single update proceeds by proposing a small change—flipping a single spin—and then accepting or rejecting this change based on the energetic cost. 

%The core logic is to \textit{always} accept a move that lowers the system's energy, but to \textit{sometimes} accept a move that increases it. This second step is crucial, as it allows the system to escape local energy minima and explore the full configuration space, mimicking the effect of a thermal bath. The acceptance probability for an energetically unfavorable move, $P(\text{accept}) = \exp(-\Delta E / T)$, ensures that at high temperatures, even costly moves are frequently accepted, while at low temperatures, the system is strongly driven towards low-energy states. A full "Monte Carlo sweep" consists of $N^2$ such attempts.

While the single-spin-flip Metropolis algorithm is conceptually simple, its performance degrades severely near the critical point. The local nature of its updates makes it extremely inefficient at altering the large-scale correlated clusters that dominate the physics at $T_c$.

To overcome this, we employ the efficient non-local Wolff cluster algorithm \cite{Wolff1989}. Instead of flipping one spin at a time, the Wolff algorithm identifies and flips an entire cluster of correlated spins in a single update. It begins by selecting a random ``seed'' spin and then recursively grows a cluster of spins by adding adjacent, like-signed neighbors with a carefully chosen probability, $P_{\text{add}} = 1 - \exp(-2J/T)$. This specific probability ensures that the collective update still satisfies the energetic balance. Once the cluster can no longer grow, all spins within it are flipped simultaneously. This allows the simulation to make large, physically relevant changes to the configuration, drastically reducing autocorrelation times and enabling efficient sampling of the equilibrium state, even at criticality. Given our focus on the behavior at and around $T_c$, the Wolff algorithm is essential for obtaining reliable results on reasonable timescales. 

In practice, we combine the two algorithms to ensure accurate thermalization. We initialize the Ising lattice in a random configuration and iteratively apply the Wolff update procedure for a large number $(>512)$ of steps. We then apply many Metropolis steps to ensure the fine-structure of the Ising configuration is realistic.

\begin{figure}[b]
  \centering
  \includegraphics[width=0.9\linewidth]{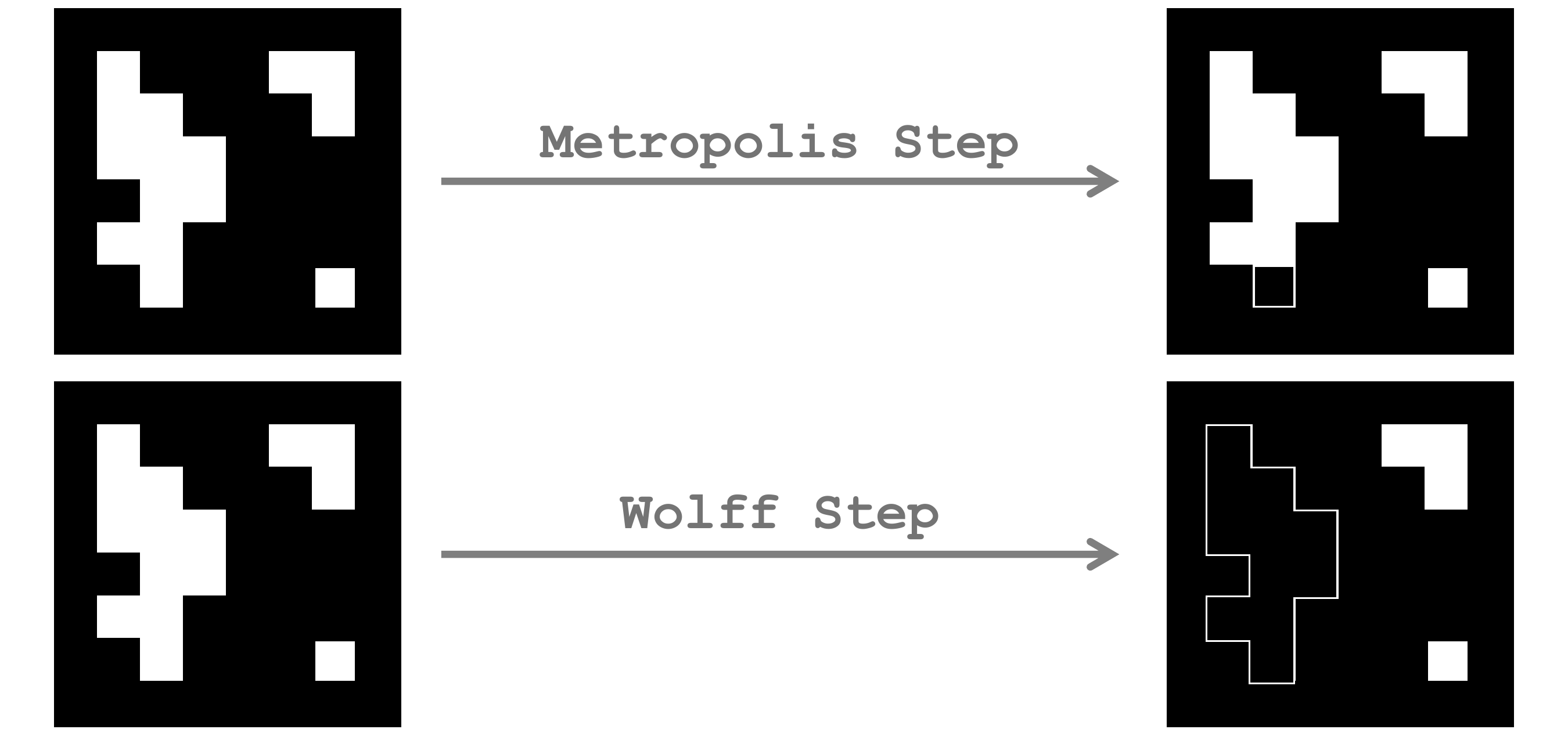}
  \caption{Visualizations of the Metropolis and Wolff lattice configuration update rules, each showing one simulation step.}
  \label{fig:Wolff_cartoon}
\end{figure}

\begin{figure}[t]
   \centering
    % You will need to generate these plots from your simulation data
   \includegraphics[width=\linewidth]{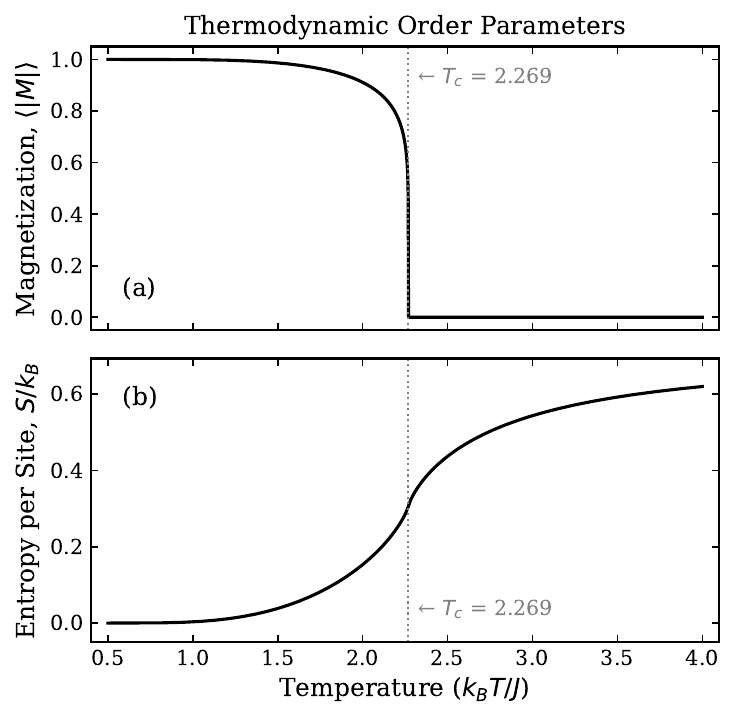}
   \caption{Behavior of traditional known standard quantities vs temperature for the 2D Ising model: (a) The average absolute magnetization, $\langle|M|\rangle$. (b) The entropy per site, $S/k_B$.}
   \label{fig:thermo_quantities}
\end{figure}

\newpage
\subsection{Relevant Thermodynamic Quantities}
\label{sec:traditional_metrics}

Before introducing our information-theoretic metric, we review the behavior of traditional physical quantities. These serve as a baseline for comparison and a validation of our simulation's accuracy. We will discuss our results in relation to two key observables: 

The \textbf{spontaneous magnetization}, $M$, is the primary order parameter for the ferromagnetic phase transition. It measures the degree of global spin alignment. Due to the up/down symmetry of the zero-field Hamiltonian, the ensemble average $\langle M \rangle$ is always zero. The physical order parameter is instead the average of the absolute magnetization, $\langle|M|\rangle = \langle |\frac{1}{N^2}\sum_i s_i| \rangle$. As shown in Fig.~\ref{fig:thermo_quantities}(a), $\langle|M|\rangle$ is non-zero for $T < T_c$, reflecting the spontaneous symmetry breaking, and falls continuously to zero at $T_c$, where the system loses its long-range order. 

%Our simulation results show excellent agreement with the exact analytical solution derived by Yang \cite{Yang1952}.

The \textbf{thermodynamic entropy}, $S$, quantifies the statistical disorder of the system, related to the number of accessible microstates via $S = k_B \ln W$. As seen in Fig.~\ref{fig:thermo_quantities}(b), the entropy is a monotonically increasing function of temperature. It is low in the ordered state, increases most steeply around $T_c$ as new configurations become accessible, and saturates at its maximum value in the high-temperature random state.

Neither of these traditional metrics peaks at the critical point. Magnetization vanishes, and entropy simply passes through an inflection point. We seek a complementary metric that would trace neither order nor disorder, but instead capture the structural richness that is maximal at their transition.

\newpage
\section{Measuring Complexity}
\label{sec:quantifying_complexity}

The challenge of quantifying the intuitive notion of complexity has led to a diverse family of theoretical proposals. Standard parameters like entropy grow monotonically with temperature, failing to capture the structural richness peaked at a transition. This has motivated the development of metrics designed to be low for both perfect order and perfect randomness. Prominent examples include ``effective complexity" \cite{GellMann1996}, ``thermodynamic depth" \cite{Lloyd1988}, and ``statistical complexity" \cite{Crutchfield1989}. While informative and interesting theoretically, these measures necessarily invoke assumptions about the best statistical model for the data source or its evolutionary history, and are thus undefined or difficult to compute for arbitrary experimental data generated by an unknown source. In contrast to these, we seek a metric that captures the same intuitive notion but which is fundamentally non-generative and ahistorical and can thus be applied to any arbitrary, multidimensional data.

\subsection{Algorithmic Complexity}

To accomplish this, we will define our metric to reflect the \textit{density} of unique patterns present in a given signal. We choose to quantify this in terms of the Kolmogorov complexity or algorithmic compressibility of the signal. The Kolmogorov complexity, $K(S)$, of a discrete signal $S$ (such as a string or an image) is defined to be the length of the shortest possible computer program which can be run on a universal Turing machine that can generate $S$ as its output and then halt \cite{Kolmogorov1965}. This measure is absolute in the sense that it is independent of the choice of programming language or machine, up to an additive constant. It represents the ultimate limit of lossless compression, capturing any and all patterns, symmetries, structures, and regularities within the given signal.

However, as a consequence of the halting problem\footnote{Alan Turing's famous halting problem entails that no general algorithm can determine whether a given arbitrary program will ever finish its execution \cite{Turing1936}. To calculate $K(S)$ exactly, one must essentially check all possible programs up to a certain length to determine if any of them produce the given signal. This is impossible because some of these programs may never halt. }, $K(S)$ is formally uncomputable. Despite this limitation, a powerful practical approach exists: $K(S)$ can be upper-bounded by the length of the compressed output, $|K_{\texttt{alg}}(S)|$, produced by any lossless compression algorithm, $\texttt{alg}$. The pioneering work of Lempel and Ziv led to practical algorithms (like LZ77) that form the basis of many modern compressors (e.g., Gzip, PNG) and whose performance is asymptotically optimal for ergodic sources \cite{Ziv1977}. This principle allows us to use standard, off-the-shelf compressors as a metric that approximates the true descriptive complexity of a given signal \cite{Kaspar1987}.

\newpage
\subsection{Compressibility as a Practical Metric}

\begin{figure}[b]
  \centering
  \includegraphics[width=0.9\linewidth]{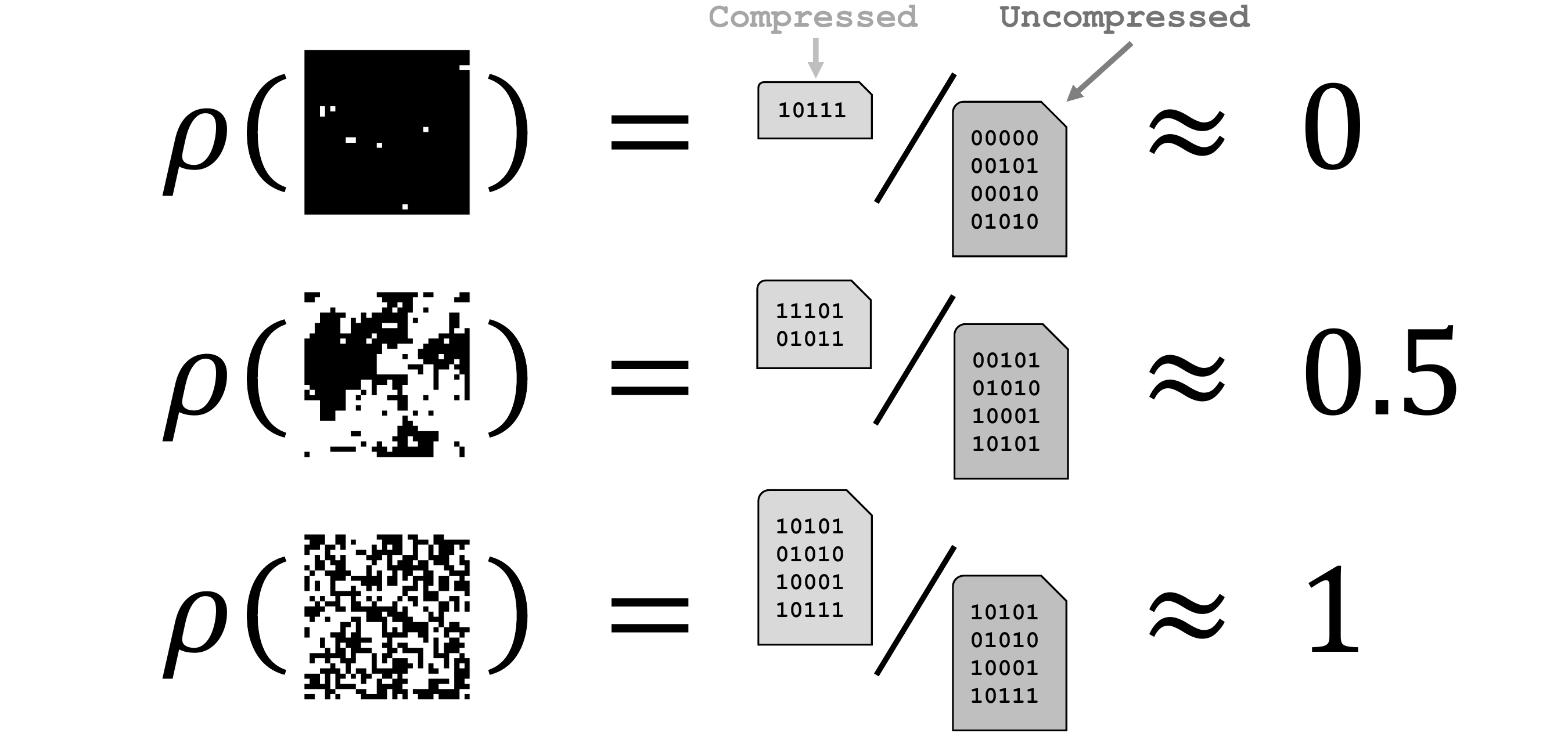}
  \caption{Visual schematic of the compression ratio, $\rho$.}
  \label{fig:rho_cartoon}
\end{figure}

To apply this concept to the Ising model, we translate a 2D Ising spin configuration into a digital image of size $N \times N$, where spin-up ($+1$) and spin-down ($-1$) are mapped to distinct pixel values (e.g., white and black). We then measure its compressed size using the commonly used lossless Portable Network Graphics (PNG) format. PNG employs the DEFLATE algorithm, a combination of LZ77 compression and Huffman coding, making it a robust estimator of algorithmic complexity for 2D data which approaches the theoretical limit of compression.

Following \cite{Martiniani2019}, we define the ``compressibility ratio," or \textit{computable information density}, $\rho[S]$, of a signal or micro-state $S$ as the ratio of its compressed size in bytes, $|K_{\texttt{alg}}(S)|$, to its original uncompressed size, $L(S)$:
\begin{equation}
    \rho[S] = \frac{|K_{\texttt{alg}}(S)|}{L(S)}.
    \label{eq:rho_def}
\end{equation}
Intuitively, a configuration with few or simple patterns (e.g., the fully ordered ferromagnetic state) will be highly compressible, yielding $\rho \to 0$. Conversely, a completely random, uncorrelated configuration (approximated by the infinite-temperature state) contains no discernible patterns and will be virtually incompressible, requiring nearly as much information to express the compressed format as the original, yielding $\rho \to 1$.

In practice, for a finite discretized signal $S$ and an imperfect compression algorithm, measurements of $\rho[S]$ may include some noise due to the finite randomness of the realized signal. However, in the infinite limit, it is expected that $\rho$ approaches a smooth monotonic function.

Numerical measurements of $\rho[S]$ for a range of Ising configurations are plotted in figure \ref{fig:complexity_vs_T}, below. For the simulated configurations, the compressibility ratio $\rho[S]$ smoothly transitions from $0 \to 1$ as a function of temperature. As one might expect, much like the thermodynamic entropy, the derivative of the compressibility ($d\rho/dT$) has a maximum near the critical temperature $T_c$, although perhaps slightly offset. 

While the behavior of $\rho$ is similar to that of entropy as a function of temperature, it is worth noting that entropy is a statistical measure of the macrostate (ie. the total Energetic content of the lattice), which by definition does not reflect any structural details of the specific signal.

\begin{figure*}
  \centering
  \includegraphics[width=\textwidth]{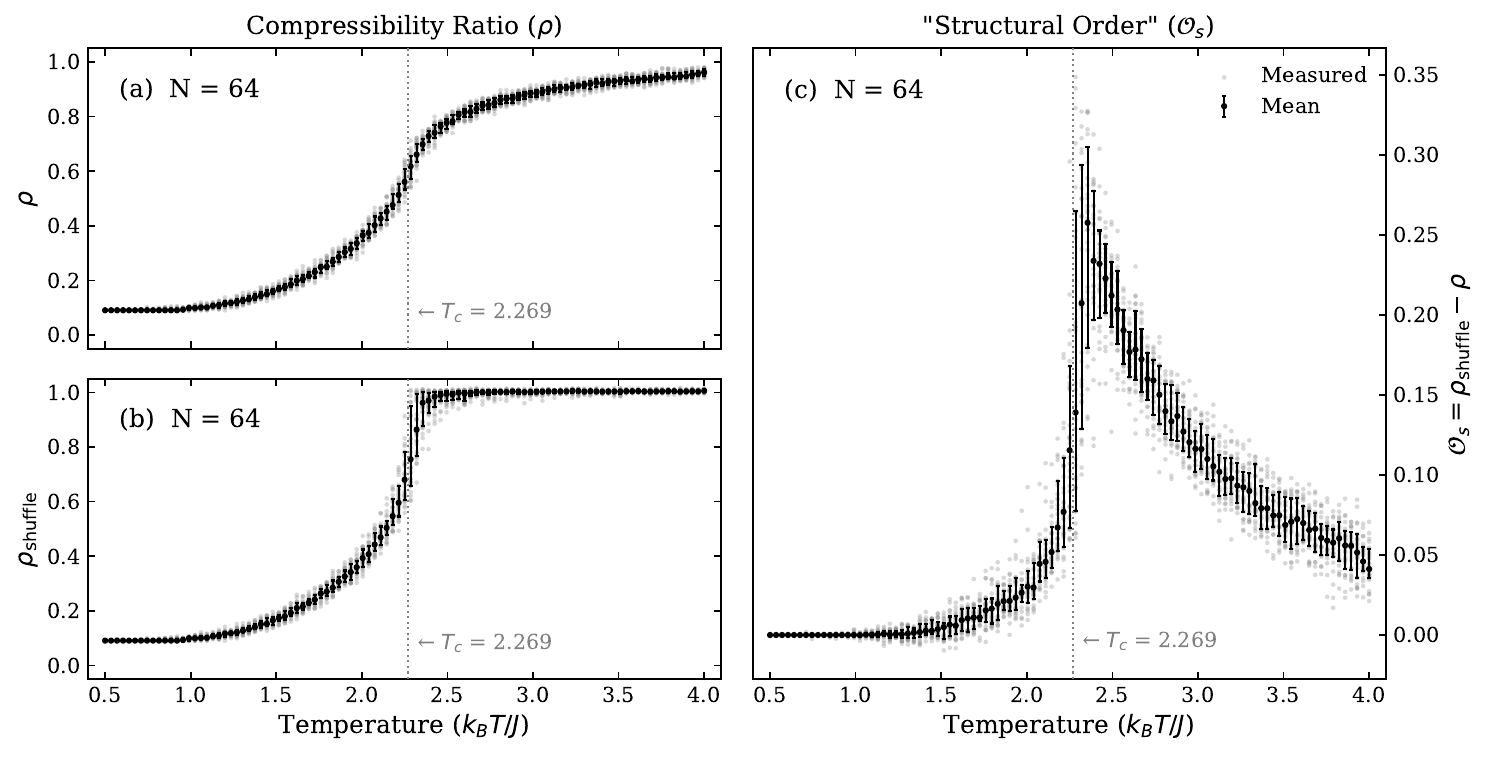}
  \vspace{-12pt}
  \caption{Result of numerical experiment measuring $\mathcal O_s$ using PNG compression. (a) shows the mean compressibility of the simulated lattices as a function of T. (b) shows the same for the shuffled lattices. (c) plots the combined metric, $\mathcal O_s$ vs T.}
  \label{fig:order_vs_T}
\end{figure*}

\newpage

\subsection{Structural Order ($\mathcal O_s$)}
\vspace{-2pt}

The compressibility $\rho$ alone is an insufficient measure of the complexity of the structure of the image because it is highly sensitive to the overall macroscopic composition of the system. For example, a low-temperature state with 99\% spins up is highly compressible simply due to this extreme statistical bias, not necessarily because of any meaningful spatial order. In the opposite limit, a high-temperature state with almost no local correlation will be nearly incompressible, despite containing virtually no coherent patterns. To isolate the contribution of spatial structure, we must normalize against a baseline that has the same composition but no spatial correlations.

Following the methodology of Melchert and Hartmann \cite{Melchert2015}, we introduce such a baseline by randomly shuffling the spins of the configuration $S$ to create a new lattice configuration $\texttt{shuffle}(S)$. This process destroys all spatial correlations while exactly preserving the number of up and down spins (i.e., the magnetization). We then define the compressibility of this baseline as $\rho_{\text{shuffle}}$:
\begin{equation}
    \rho_{\text{shuffle}}[S] = \rho[\texttt{shuffle}(S)].
    \label{eq:rho_shuffle_def}
\end{equation}
This normalizing term $\rho_{\text{shuffle}}$ represents the compressibility of a related configuration that is as random as possible given its composition, some microstate which is exactly consistent with the macrostate but otherwise contains no internal structure. In general, in the infinite limit any subtle spatial structure contained by the signal will allow an ideal compressor to extract some reducible information, therefore: $\rho[S] \leq \rho_{\text{shuffle}}[S]$.

\newpage

The normalized difference between $\rho$ and $\rho_{\text{shuffle}}$ reveals the ``compression bonus'' gained from the presence of any spatial structure of the original lattice. To quantify this, we define the term ``Structural Order,'' $\mathcal O_s$, as:

\vspace{-5pt}

\begin{equation}
    \mathcal O_s[S] = \rho_{\text{shuffle}}[S] - \rho[S] %=  \frac{|K_{\texttt{alg}}(S)| - |K_{\texttt{alg}}(\texttt{shuffle}(S))|}{N^2}.
    \label{eq:Os_def}
\end{equation}

\vspace{5pt}

This metric represents the fractional improvement in compression due to the presence of non-trivial spatial correlations. If a configuration has no spatial structure beyond random chance, $\rho[S] \approx \rho_{\text{shuffle}}[S]$, and thus $\mathcal  O_s \to 0$. If a configuration has significant spatial order that is destroyed by shuffling, then $\rho[S] \ll \rho_{\text{shuffle}}[S]$, causing $\mathcal O_s$ to approach its maximum value. We therefore expect that $\mathcal O_s$ will be near zero in both the highly ordered and highly disordered regimes, and will exhibit a maximum at the critical temperature $T_c$, where long-range structural correlations are most prominent (see Fig.~\ref{fig:order_vs_T} above).

\begin{figure}[b]
  \centering
  \includegraphics[width=0.9\linewidth]{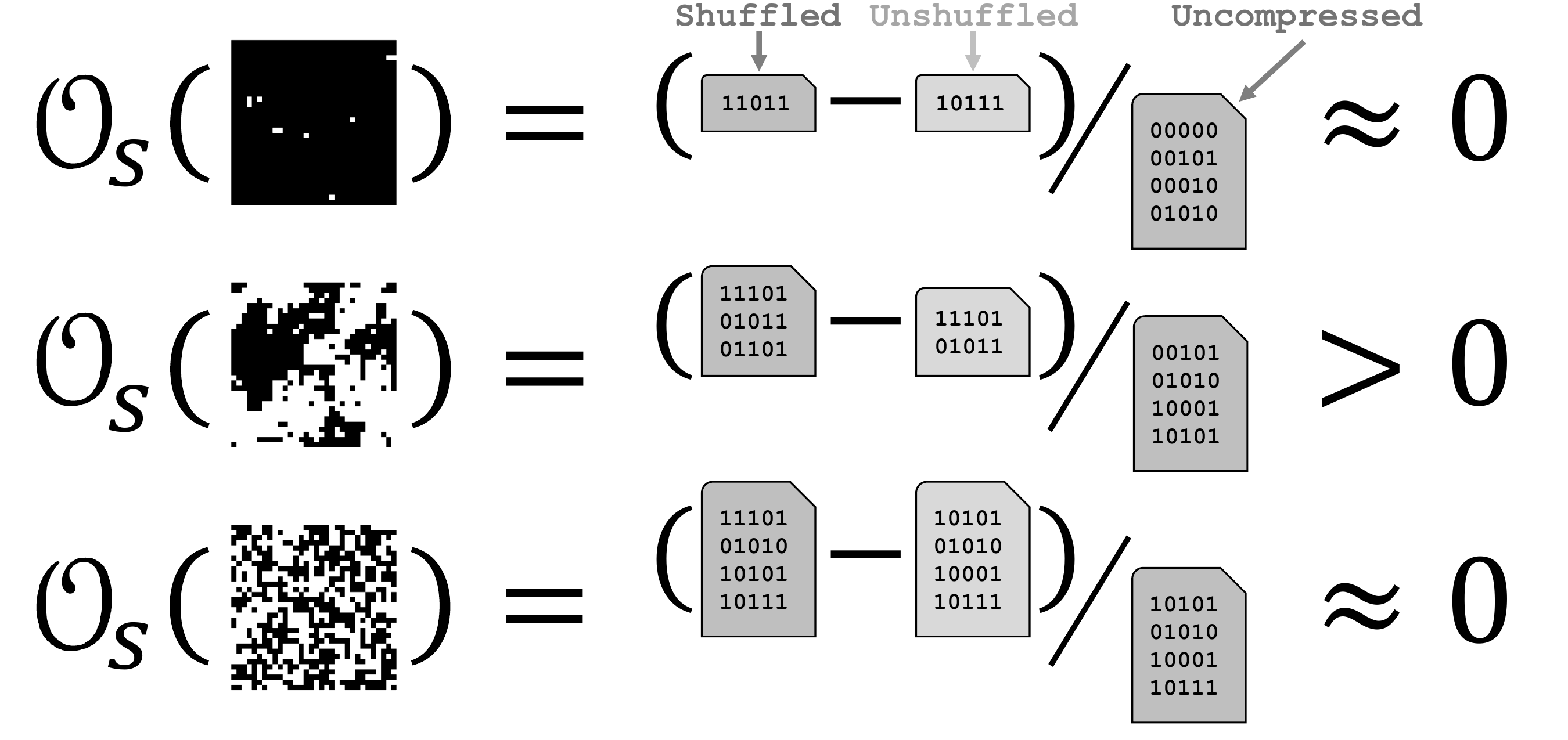}
  \caption{Visual schematic of the Structural Order metric, $\mathcal O_s$.}
  \label{fig:Os_cartoon}
\end{figure}

\begin{figure*}
  \centering
  \includegraphics[width=\textwidth]{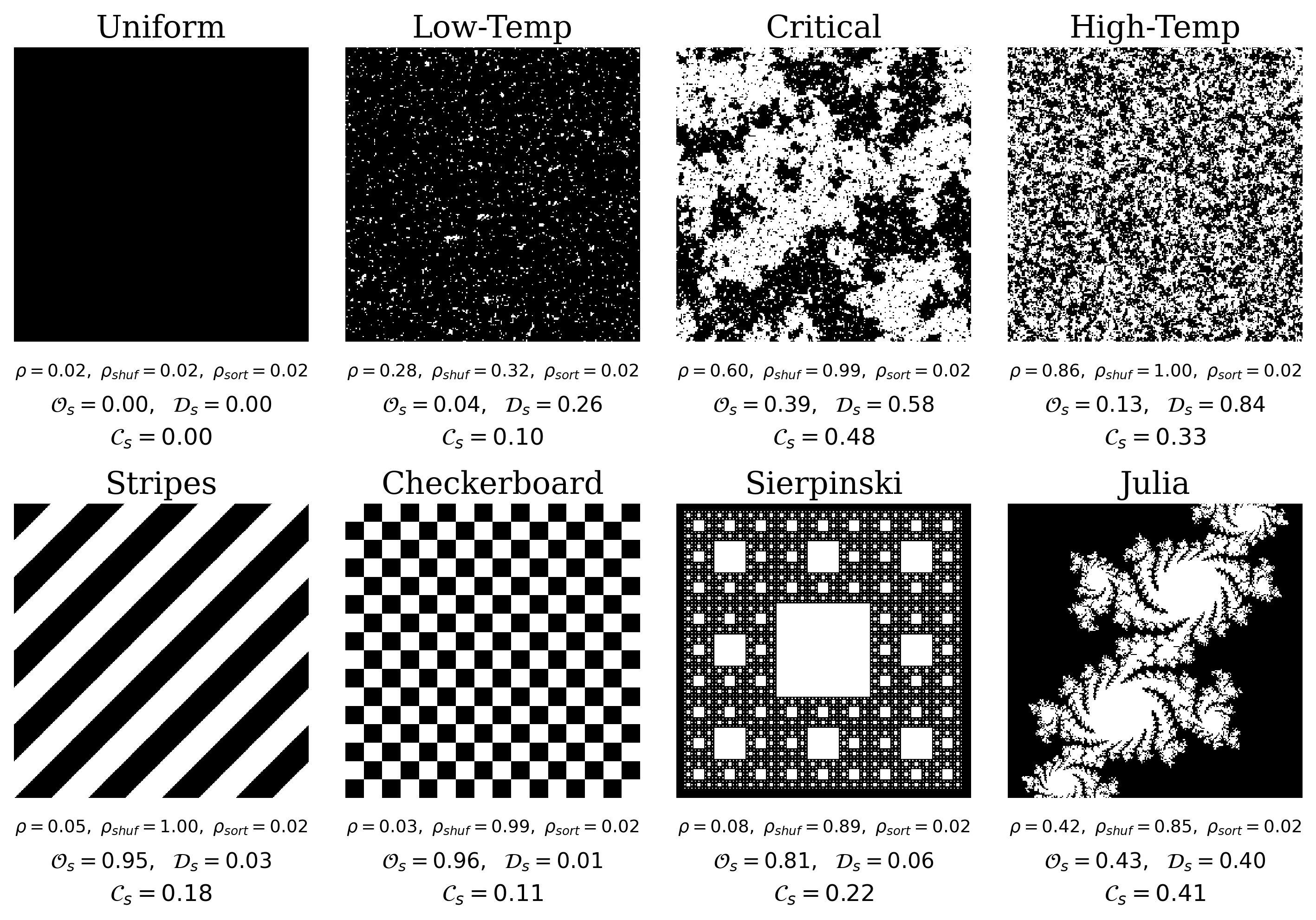}
  \vspace{-12pt}
  \caption{Benchmark examples of the compressibility parameters $\rho$, $\rho_{\text{shuffle}}$, and $\rho_{\text{sort}}$ and the combined structural content metrics $\mathcal{O}_s = (\rho_{\text{shuffle}} - \rho)$, $\mathcal{D}_s = (\rho - \rho_{\text{sort}})$ and $\mathcal C_s = \sqrt{\mathcal O_s \times \mathcal D_s}$. Top row shows sample simulated Ising lattices of increasing temperature. Bottom row shows ``out of distribution'' configurations meant to examine the behavior of these metrics.}
  \label{fig:complexity_metric_showcase}
\end{figure*}

\newpage
\subsection{The Challenge of Further Extracting ``Interesting'' or ``Aperiodic'' Structure}

We examine the behavior of the compressibility parameters, $\rho$, and the structural order metric, $\mathcal O_s$, on several benchmark configurations in figure \ref{fig:complexity_metric_showcase} above.
We notice that the structural order metric, which we quantify as $\mathcal{O}_s = (\rho_{\text{shuffle}} - \rho)$, identifies both simple periodic patterns (ie. a checkerboard) and complex fractal patterns (ie. the critical state) as being highly ordered, while random configurations are assigned low values. This suggests that this term $\mathcal O_s$ can meaningfully quantify a system's departure from structural randomness. 
However, while this parameter is sufficient to distinguish the critical phase transition in the Ising model (compared to other Ising states), it does not meaningfully capture an intuitive notion of ``complexity." The behavior of $\mathcal O_s$ for the Ising model as a function of Temperature (as shown in Fig.~\ref{fig:order_vs_T}) \textit{does} indeed behave how we intuitively expect a metric of complexity should, being low for both low and high-temperature states with a peak around $T \approx T_c$.
However, the more uniform low-temperature states are assigned a low value of $\mathcal O_s$, not because they are truly unstructured or disordered, but because their uniform \textit{composition} (nearly all spins aligned) means that the shuffling operation cannot destroy any of their uniform structure, thus $\rho_{\text{shuffle}} \approx \rho$ and $\mathcal O_s \rightarrow 0$.

This failure highlights a broader fundamental challenge in complexity science: the need to distinguish between the simple, ``crystalline" regularity of a periodic lattice and the sophisticated, fractal structure of a critical system \cite{Crutchfield2012}. While both are non-random, one would not describe a simple periodic pattern as ``complex" in the same vein as the hierarchical structures characteristic of systems like the critical Ising state or a living organism.

This observation motivates an extension of our structural order metric to better distinguish between simple, crystalline order and more complex, fractal structure. A truly meaningful metric of ``complexity'' should be low for both absolute randomness and periodic order. To achieve this, we introduce a ``structural disorder'' metric which will quantify a given state's departure from simplicity.

\begin{figure*}
  \centering
  \includegraphics[width=\textwidth]{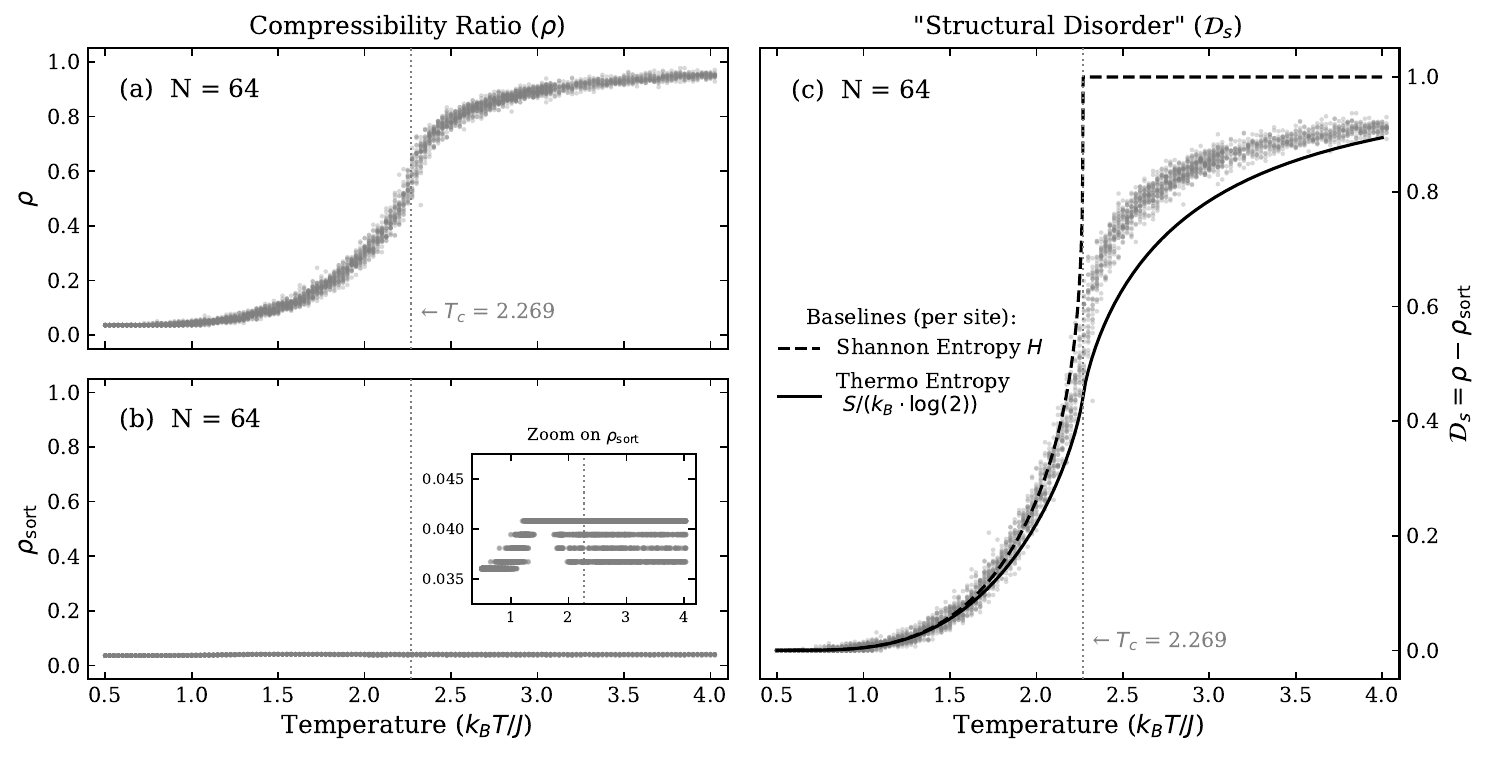}
  \vspace{-12pt}
  \caption{Result of numerical experiment measuring $\mathcal D_s$ using PNG compression. (a) shows the mean compressibility of the simulated lattices as a function of T. (b) shows the same for the sorted lattices, with a zoom to visualize the dynamic range. (c) shows $\mathcal D_s$ compared with the Thermodynamic and Shannon Entropies per site, $S$ \& $H$, normalized to the range [0,1]. }
  \label{fig:disorder_vs_T}
\end{figure*}

\newpage

\subsection{Structural Disorder ($\mathcal D_s$)}

To address the challenge of meaningfully quantifying the structural complexity of a given image, we introduce a new, complementary metric which we term ``Structural Disorder," denoted $\mathcal{D}_s$. The purpose of $\mathcal{D}_s$ is to quantify how far a configuration's spatial arrangement is from its simplest possible, most ordered state, given a fixed composition of spins. 

To measure a departure from simplicity, we must first establish a baseline state of maximal simplicity. For any given configuration $S$ with a fixed number of up and down spins, the simplest possible spatial arrangement is one that \textit{minimizes} the descriptive length of the boundaries between separate regions. This state is achieved by grouping all up-spins into one contiguous block and all down-spins into another, creating a single domain wall between them. We will refer to this new lattice as the ``sorted" configuration, $\texttt{sort}(S)$.

Operationally, we construct $\texttt{sort}(S)$ by flattening the spins of a given lattice, $S$, into a 1-dimensional string, sorting this string by value (e.g., all `+1`s, then all `-1`s), and reshaping it back into a lattice of the original dimensions. We then calculate its compressibility, $\rho_{\text{sort}}[S]$, using the same compression algorithm as before. This value represents the algorithmic complexity of the most ordered possible state for that specific spin composition and lattice size. For any non-trivial composition, $\rho_{\text{sort}}$ will be small but non-zero, as the compressor must still encode the boundary between the two spin domains.

\newpage 

With this baseline established, we define the Structural Disorder, $\mathcal{D}_s$, as the normalized difference between the compressibility of the original configuration and that of its sorted counterpart:

\vspace{-5pt}

\begin{equation}
    \mathcal D_s[S] = \rho[S] - \rho_{\text{sort}}[S] %=  \frac{|K_{\texttt{alg}}(S)| - |K_{\texttt{alg}}(\texttt{shuffle}(S))|}{N^2}.
    \label{eq:Ds_def}
\end{equation}

\vspace{5pt}

This metric quantifies the ``compressibility cost" from arranging the spins into their actual spatial pattern, compared to the simplest possible arrangement. $\mathcal D_s$ is characteristically low for simple, periodic structures like the checkerboard (see Fig.~\ref{fig:complexity_metric_showcase}) which are already nearly maximally compressible. Figure \ref{fig:disorder_vs_T} above plots the behavior of $\mathcal D_s$ vs temperature for our 2D Ising model, and shows that it has a very similar form to the Entropy per site, S, a well-established metric for disorder calculated from the macrostate.

\begin{figure}[b]
  \centering
  \includegraphics[width=0.9\linewidth]{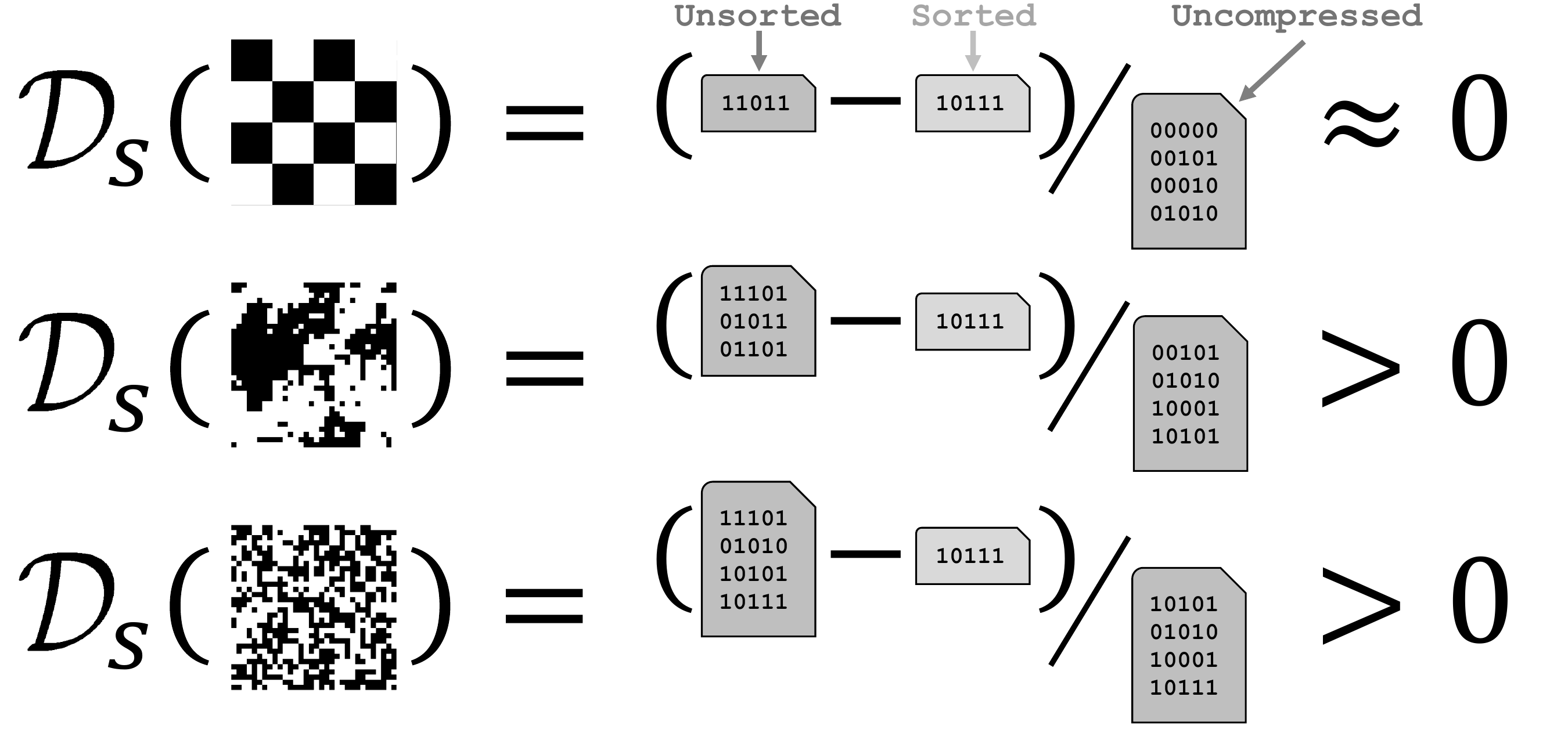}
  \caption{Visual schematic of the Structural Disorder metric.}
  \label{fig:Ds_cartoon}
\end{figure}

\begin{figure*}
  \centering
  \includegraphics[width=\textwidth]{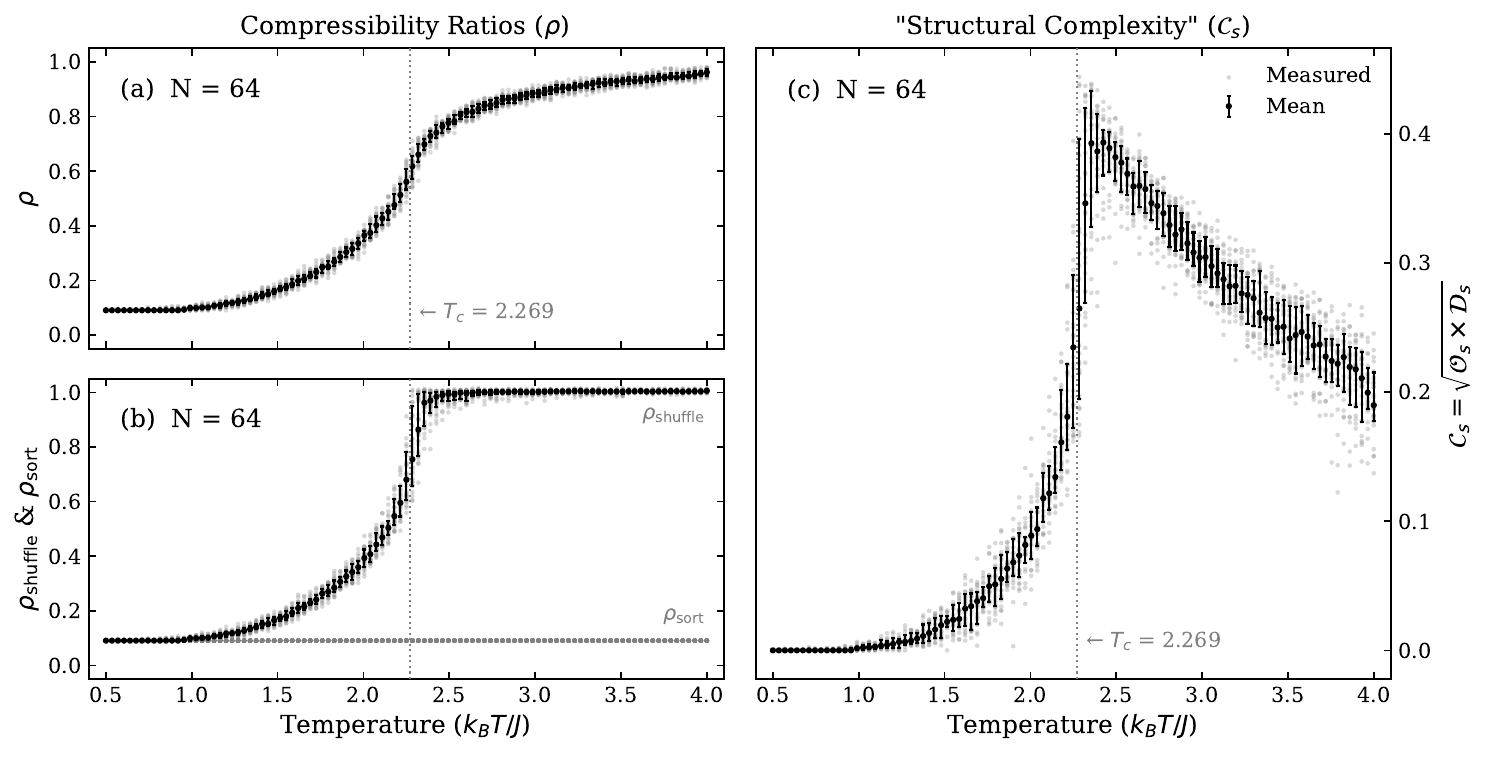}
  \vspace{-12pt}
  \caption{Result of numerical experiment measuring $\mathcal C_s$ using PNG compression. (a) shows the mean compressibility of the simulated lattices as a function of T. (b) shows the same for the shuffled lattices. (c) plots the combined metric, $\mathcal C_s$ vs T.}
  \label{fig:complexity_vs_T}
\end{figure*}

\newpage

\subsection{Structural Complexity ($\mathcal C_s$)}

In the preceding sections, we introduced two complementary, information-theoretic metrics designed to probe the spatial structure of a physical system, given a snapshot of its microstate. The first metric, ``Structural Order'' ($\mathcal{O}_s$), quantifies a configuration's departure from statistical randomness. The second, ``Structural Disorder'' ($\mathcal{D}_s$), quantifies its departure from trivial, perfectly sorted order. We have established that, while $\mathcal{O}_s$ is an effective ``regularity detector,'' it is insufficient to fully capture an intuitive notion of complexity on its own. As demonstrated in figure \ref{fig:complexity_metric_showcase}, $\mathcal{O}_s$ assigns a high value to simple periodic patterns, failing to distinguish them from the hierarchical structures characteristic of true, ``emergent'' complexity.

This distinction is central to complexity science. True ``emergent'' complexity, as seen in critical phenomena, biological organisms, and other adaptive systems, is not merely non-random; it occupies a specific, rarefied space between the two extremes of featureless randomness and rigid, crystalline order \cite{Crutchfield2012}. A robust complexity metric should therefore be vanishing for both limiting cases.

To formalize this requirement, we propose a new ``Structural Complexity'' metric, which we will henceforth refer to as $C_s$, constructed from our two previously defined components. The central assumption is that sophisticated structure can only exist in a system that is simultaneously highly ordered (far from random) \textit{and} highly disorganized (far from simple), and which should therefore have large values in both $\mathcal{O}_s$ and $\mathcal{D}_s$ metrics.

The most natural mathematical form of this logical \texttt{AND} condition is a multiplicative expression. We therefore choose to define our final structural complexity metric, $\mathcal C_s$, as the geometric mean of the Structural Order and Structural Disorder metrics:
\begin{equation}
    \mathcal C_s = \sqrt{\mathcal{O}_s \times \mathcal{D}_s} = \sqrt{(\rho_{\text{shuffle}} - \rho) \times (\rho - \rho_{\text{sort}})}
    \label{eq:cs_final_def}
\end{equation}

This formulation has several powerful and desirable properties. The multiplicative nature ensures that $C_s$ can only be significantly non-zero if both of its constituent terms, $\mathcal{O}_s$ and $\mathcal{D}_s$, are large. If a system is random, its $\mathcal{O}_s \approx 0$, forcing $C_s \to 0$. If a system has simple, periodic order, its $\mathcal{D}_s \approx 0$, which also forces $C_s \to 0$. A large $C_s$ value is thus reserved exclusively for systems that are rich in non-random, aperiodic structure. 
%$\mathcal{C}_s$ can thus be interpreted as a measure of a system's ``structural sophistication" or ``aperiodic order." 
We adopt a square root to map the product back to the linear scale of compressibility. Due to the $[0,1]$ range of the compressibilities, $\mathcal{C}_s$ also has the useful property of being upper-bounded by a known maximum $\mathcal{C}_s \leq \mathcal{C}_s^*(S) \leq 1/2$ for a given composition (i.e., fixed $\rho_{\text{sort}}$ and $\rho_{\text{shuffle}}$) where $\mathcal{C}_s^*(S) = (\rho_{\text{shuffle}}(S) - \rho_{\text{sort}}(S))/2$. The metric $\mathcal{C}_s$ also naturally respects key symmetries of the system, being invariant under translation, rotation, reflection, and even pixel-wise inversion (subject to choice of $\texttt{alg}$). 

This construction is a practical, data-driven realization of a deep and well-established principle in the study of complexity, which holds that complexity can only exist somewhere between perfect order and disorder (see, for instance, Fig. 2 of \cite{Huberman1986}). Our ``Structural Complexity'' metric $\mathcal{C}_s$ directly inherits from this concept and further extends it to static image processing.

%By employing this refined metric, we expect to generate a complexity spectrum, $C_s(L,T)$, that is truly selective for the emergent, hierarchical structures characteristic of criticality. The remainder of this work will utilize this second-order definition of $C_s$ to analyze the complexity landscape of the 2D Ising model.

%This approach is conceptually aligned with the ``thermodynamic depth" proposed by Lloyd and Pagels, which equates complexity with the amount of information processing needed to generate a state from a simple origin \cite{Lloyd1988}. Here, $\mathcal{D}_s$ serves as a practical, static proxy for this idea, measuring the descriptive ``distance" of a state from its simplest possible sorted configuration. By combining this with our measure of Structural Order, $\mathcal{O}_s$, we can construct a new, second-order complexity metric, $C_s = \sqrt{\mathcal{O}_s \times \mathcal{D}_s}$, which is large only for states that are simultaneously far from random and far from simple—the very definition of emergent complexity.

\newpage

\section{Numerical Results}
\label{sec:results}

\begin{figure*}
  \centering
  \includegraphics[width=\textwidth]{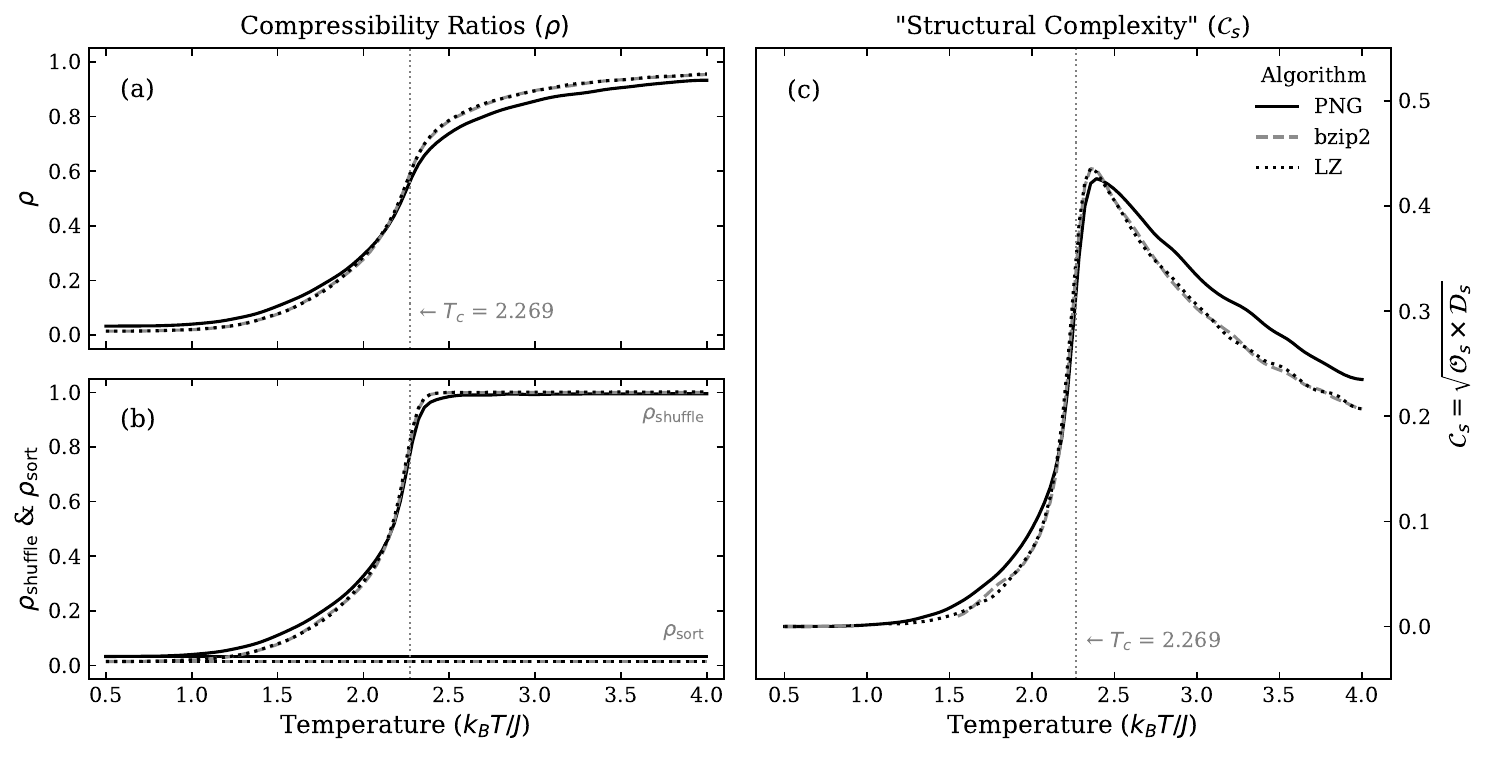}
  \vspace{-12pt}
  \caption{Result of numerical experiments measuring $\mathcal C_s$ using various compression algorithms. Plots show a Gaussian-smoothed mean of the simulated data to highlight the broader trend. (a) and (b) show the mean compressibility of the simulated lattices and their shuffled and sorted counterparts, respectively, as a function of T. (c) plots the combined metric, $\mathcal C_s$ vs T. All curves show the metrics for Ising lattice size $N=128$.}
  \label{fig:alg_comp}
\end{figure*}

\begin{figure*}
  \centering
  \includegraphics[width=\textwidth]{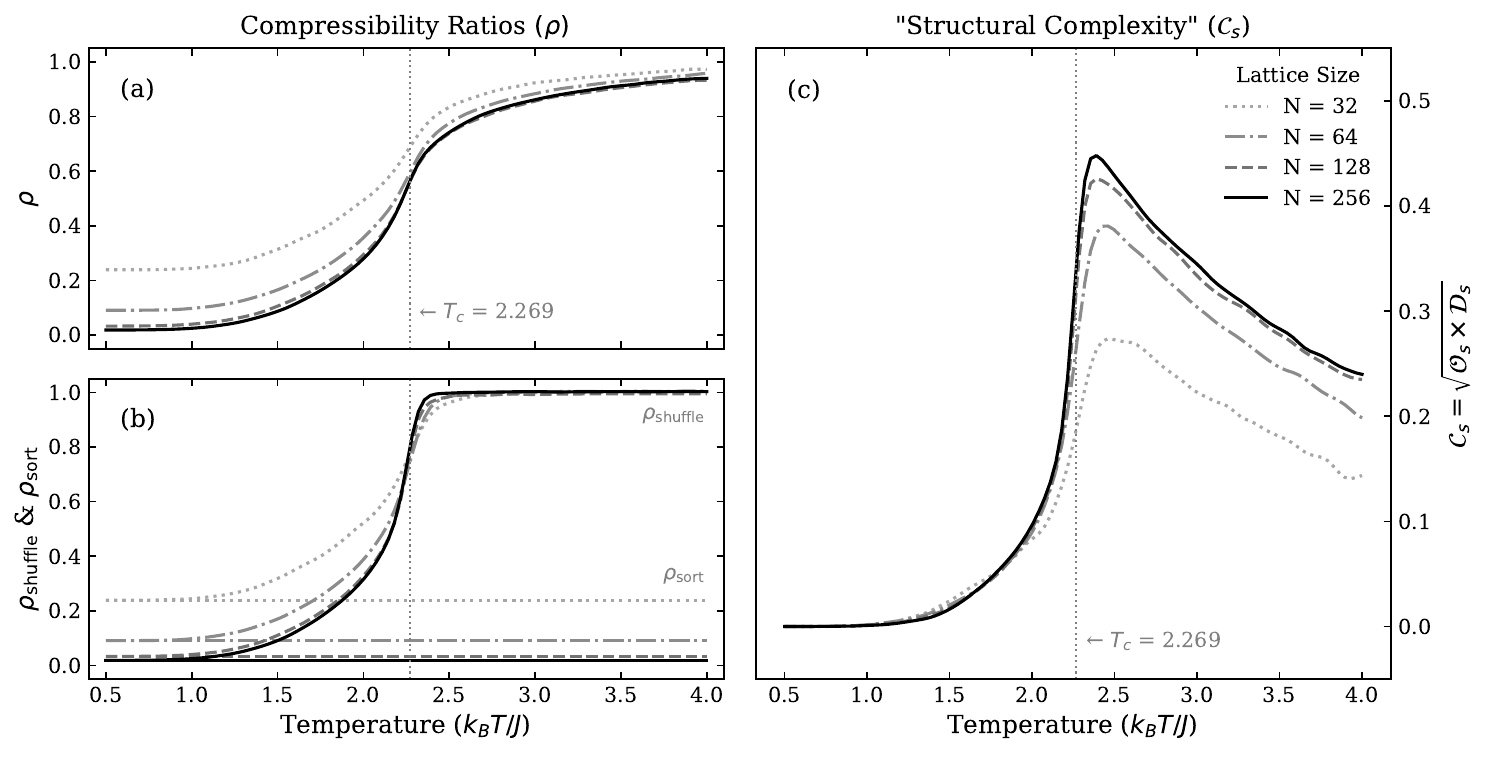}
  \vspace{-12pt}
  \caption{Result of numerical experiments measuring $\mathcal C_s$ using PNG compression for a range of finite lattice sizes N. Plots show a Gaussian-smoothed mean of the simulated data to highlight the broader trend. (a) and (b) show the mean compressibility of the simulated lattices and their shuffled and sorted counterparts, respectively, as a function of T. (c) plots the combined metric, $\mathcal C_s$ vs T. Notice that the spike in the complexity metric $\mathcal C_s$ becomes taller and somewhat sharper for larger lattice sizes.  The raw numerical data and fits for each N are shown separately in Fig.~\ref{fig:N_comp_all} in the appendix.}
  \label{fig:N_comp}
\end{figure*}

\begin{figure*}
  \centering
  \includegraphics[width=\textwidth]{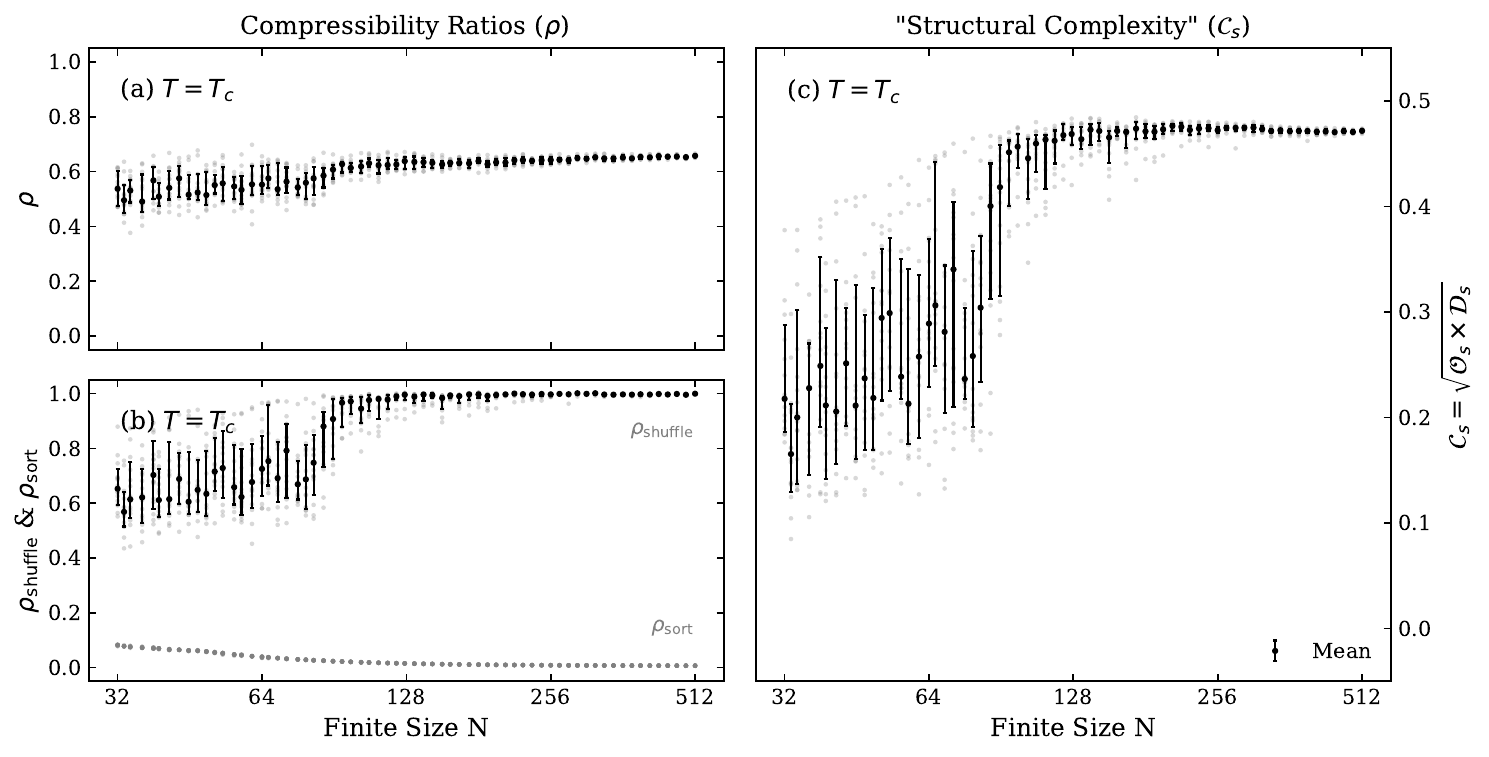}
  \vspace{-12pt}
  \caption{Result of numerical experiments measuring $\mathcal C_s$ at the critical temperature $T=T_c$ for a range of finite sizes N. Plots show individual simulations in grey and a mean and interquartile range of a given size in black. Plots (a) and (b) show the mean compressibility of the simulated lattices and their shuffled and sorted counterparts, respectively, as a function of size N. (c) plots the combined metric, $\mathcal C_s$ vs N. Note that $\mathcal C_s(T_c)$ approaches an asymptotic limit just below its theoretical maximum.}
  \label{fig:finite_N}
\end{figure*}

Having now established the context and definition of our proposed metric, in this section, we now present the results of our numerical experiments calculating $\mathcal C_s$ for simulated data and observing its behavior. We first apply our structural complexity metric, $\mathcal C_s$, to a wide range of Ising lattice configurations to see how it behaves as a function of temperature and to demonstrate its ability to identify the critical point (Fig.~\ref{fig:complexity_vs_T}). We then test the robustness of our result by varying the algorithm used to approximate the Kolmogorov complexity (Fig.~\ref{fig:alg_comp}). We examine the finite size dependence of the behavior of $\mathcal C_s$ by comparing results while varying the lattice size (Figs.~\ref{fig:N_comp}, \ref{fig:finite_N}). Finally, we explore the behavior of $\mathcal C_s$ vs T around the critical point in more depth, quantitatively, and fit a phenomenological model which allows us to tease out some interesting properties.

%We then utilize the block-spin coarse-graining technique to generate a ``structural complexity spectrum," $\mathcal C_s(L)$, which quantitatively reveals the scale-invariant nature of the critical state. We compare this to several other Ising lattice configurations and popular fractal images (Fig.~\ref{fig:Unique_Lattice_Spectra}) to better examine the behavior of $\mathcal C_s(L)$ and its applicability as a detector of self-similarity. Finally we generate a heatmap showing the Complexity Spectra $\mathcal C_s(L)$ of the Ising lattices for a range of temperatures (Fig.~\ref{fig:ising_block_spectrogram}), to better examine the scale-dependence of its behavior around the phase transition. 

All simulations of the 2D Ising model are performed on square lattices of with periodic boundary conditions, using the efficient Wolff cluster algorithm to accelerate convergence around the critical point \cite{Wolff1989}, and the slower Metropolis algorithm to ensure thermalization \cite{Metropolis1953}.  We perform a temperature sweep from $T=0.5$ to $T=4.0$ in steps of $\Delta T=0.035$. For each temperature, we generate at least $16$ lattices from unique initial conditions to suppress any statistical fluctuations in our measurements.

The complexity metrics are calculated using a lossless compression algorithm (PNG, unless otherwise stated). The values of $\rho$, $\rho_{\text{sort}}$, and $\rho_{\text{shuffle}}$ are calculated by comparing the raw compressed byte-count of the target lattice and its randomly shuffled version to the uncompressed lattice size and are thus presented in dimensionless units. To account for any preference the compression algorithm may have for lighter or darker pixels, we calculate the compressibilities, $\rho$, for both a given lattice and its inverted counterpart and record their average.
To suppress random noise in the shuffling procedure, we independently shuffle the lattice $16$ times and let $\rho_{\text{shuffle}}$ be the average compression ratio of this ensemble. 
The structural complexity is then computed as $\mathcal C_s = \sqrt{\mathcal{O}_s \times \mathcal{D}_s} = \sqrt{(\rho_{\text{shuffle}} - \rho) \times (\rho - \rho_{\text{sort}})}$, where the compressibility values are averaged as described before the calculation of $\mathcal C_s$.

\vspace{14pt}

\hrule

\vspace{14pt}

The proposed ``structural complexity" metric ($\mathcal C_s$) and its component factors, ``structural order" ($\mathcal O_s$) and ``structural disorder'' ($\mathcal D_s$), are plotted vs temperature on the proceeding pages for a densely-sampled suite of simulations of finite size $N = 64$. From these alone, it is empirically clear that $\mathcal C_s$ is strongly sensitive to the critical phase transition and has a peak value at or around the critical point. The following subsections explore the behavior of $\mathcal C_s$ across parameter space in greater depth.

\newpage
\subsection{Choice of Compression Algorithm}

A central premise of our work is that the structural complexity peak at $T_c$ is a fundamental property of the system's information content, not an artifact of a specific compression algorithm. To validate this, we performed the same analysis using several different lossless compression algorithms, each based on distinct underlying principles. We benchmarked our primary method, PNG (which uses the DEFLATE algorithm, a combination of LZ77 and Huffman coding), against two standard 1D compression schemes: LZMA (Lempel-Ziv-Markov chain algorithm) and bzip2 (which uses the Burrows-Wheeler Transform). For the 1D algorithms, the 2D lattice configurations were flattened into a 1D byte string using a standard row-major scanline order before compression.

The results, shown in Fig.~\ref{fig:alg_comp}, compellingly demonstrate the robustness of our findings. All three algorithms produce a structural complexity curve, $C_s(T)$, that exhibits a sharp, unambiguous peak precisely at the theoretical critical temperature, $T_c \approx 2.269$. This confirms that the presence of emergent complexity at the phase transition is detectable regardless of the algorithm used.

However, we observe interesting quantitative differences between the algorithms. The peaks produced by the 1D algorithms, LZMA and bzip2, are noticeably thinner and sharper than the peak produced by the 2D-aware PNG algorithm. This is likely because the 1D methods are structurally ``blind" to regularity along the vertical axis. Their ability to find correlations is limited to patterns that occur along the 1D horizontal scanline. Around criticality, the Ising model's domains are large, 2D isotropic clusters. A 2D algorithm like PNG, which can employ filters that look at vertical neighbors, is capable of discovering these larger 2D regularities. This makes its compression more efficient across a broader range of near-critical temperatures where large domains are present, resulting in a wider peak in the structural complexity metric. This effect is a result of the fact that these algorithms can only asymptotically approach the true Kolmogorov complexity, and demonstrates that some do so better than others.

Intriguingly, while the peak widths differ, the maximum value of $C_s$ at $T_c$ is remarkably similar across all three algorithms. We speculate that this is a manifestation of the scale-invariance at the critical point itself. At $T_c$, the fractal-like domain structure means that correlations exist at all length scales, including the very small scales accessible to a 1D scanline-based algorithm. While a 2D algorithm can exploit the large-scale correlations more effectively away from $T_c$, right at the critical point, the sheer density of information-rich structure at every scale may saturate the capabilities of all algorithms to a similar degree. The peak height may therefore represent a more universal measure of the maximal degree of information gain from knowing the system is structured, a value that is less dependent on the specific strategy used to discover that structure.

\newpage

\newpage

\subsection{Finite Lattice Size Scaling}

The sharp phase transition of the 2D Ising model is a feature of the thermodynamic limit ($N \to \infty$). Any numerical simulation, being performed on a finite lattice, will exhibit ``finite-size effects," where the sharp singularities of the transition are rounded and shifted. Studying how our complexity metric behaves as a function of the system size $N$ is therefore a critical test of their validity and provides deeper insight into its physical meaning.

In Fig.~\ref{fig:N_comp}, we present a comparison of the compressibility ratios ($\rho$, $\rho_{\text{shuffle}}$, $\rho_{\text{sort}}$) and the resulting structural complexity ($C_s$) for systems of sizes: $N=32, 64, 128, 256$. The Gaussian-smoothed mean values vs Temperature are over-plotted for each N. A clear trend emerges that is consistent with established theories that describe self-similarity and critical phenomena.

Most importantly, the peak in the structural complexity $C_s(T)$ becomes progressively taller and somewhat more sharply defined as the system size $N$ increases. For the smallest lattice ($N=32$), the peak is a broader hump maximized near the theoretical $T_c$. For the largest lattice ($N=256$), it has sharpened into a pronounced spike with roughly twice the peak value. This behavior is a classic signature of a second-order phase transition, where the order parameter's susceptibility diverges in the thermodynamic limit. We have designed our metric to be non-divergent and so in our case the value at the critical point asymptotes towards a limiting maximum values.

This behavior is due to a systematic finite-size effect of the baseline compressibility values. At low temperatures ($T \ll T_c$), the minimum values of both $\rho(T)$ and $\rho_{\text{shuffle}}(T)$ are noticeably larger for smaller lattices. A smaller system necessarily has fewer opportunities for repeated patterns and the compressed file size must also include a short header which is a larger fraction of the compressed size for smaller lattices. 
Likewise, even for a very high temperature configuration, a small random lattice will by chance have some short regularities which a compression algorithm will be able to exploit to reduce the file size by some amount. 
However, as $N \to \infty$, these baseline compressibilities approach zero and one, corresponding to an infinitely compressible ordered state and a absolutely incompressible disordered state, respectively.

The behavior of the compressibility ratios $\rho(T_c)$, $\rho_{\text{shuffle}}(T_c)$, $\rho_{\text{sort}}(T_c)$, and the complexity metric $\mathcal C_s(T_c)$ at the critical temperature are shown in greater detail as a function of finite size N in Fig.~\ref{fig:finite_N} above. Recall that the maximum possible value of our complexity metric depends on the values of a given input's shuffled and sorted compressibilities ($\mathcal{C}_s^{\text{max}} = (\rho_{\text{shuffle}} - \rho_{\text{sort}})/2$) and thus the values of $\mathcal{C}_s(T_c)$ can approach a maximum value of $\frac{1}{2}$ as the gap between $\rho_{\text{shuffle}}(T_c)$ and $\rho_{\text{sort}}(T_c)$ widens to one. For small values of N, the peak values of $\mathcal C_s$ is small and fluctuates widely due to sensitivity to random noise. But for values of N at and above around 256, the critical value of $\mathcal C_s$ approaches a finite upper bound and are far less sensitive to noise due to their size.

%Finally, the shapes of the compressibility curves themselves are highly informative. The compressibility of the structured lattice, $\rho(T)$, increases monotonically in a smooth, slightly asymmetric sigmoidal fashion. Its shape is qualitatively similar to that of the thermodynamic entropy, $S(T)$, as both metrics track the system's general progression from a simple, ordered state to a complex, disordered one. The most striking relationship is seen in the shuffled compressibility, $\rho_{\text{shuffle}}(T)$, which appears to complement the spontaneous magnetization, $\langle|M(T)|\rangle$. It is low where magnetization is high ($T < T_c$) and abruptly plateaus to its maximum value exactly where magnetization vanishes ($T \geq T_c$). This is because the compressibility of a random sequence is related to its Shannon entropy, which is minimal for a highly biased composition and maximal for an equal composition. Therefore, the compositional entropy is maximal when $\langle|M(T)|\rangle=0$. Thus, $\rho_{\text{shuffle}}(T)$ effectively measures the system's compositional disorder. 

\newpage 

\subsection{Behavior around the Critical Point}

The principle result of our work is the behavior of the structural complexity, $\mathcal C_s$, as a function of temperature. As shown in Fig.~\ref{fig:complexity_vs_T}, our metric exhibits an unambiguous peak at the critical temperature and decays towards zero in the high and low-temperature limits. However, much of the sharpness and asymmetry that appears in the peak in Fig.~\ref{fig:complexity_vs_T} is due to the sudden transition in composition (ie. magnetization, see Fig.~\ref{fig:thermo_quantities}) at the critical point. To account for this when analyzing the shape of the critical peak, we normalize $\mathcal C_s$ by its \textit{composition-dependent} maximum value, $\mathcal{C}_s^{*} \propto (\rho_{\text{shuffle}} - \rho_{\text{sort}})$. The normalization parameter $\mathcal{C}_s^{*}$ is dominated by the behavior of $\rho_{\text{shuffle}}$, which, as we have established, acts as a proxy for the system's compositional disorder, undergoing a steep, step-like transition at $T_c$ that mirrors the onset of spontaneous magnetization. The normalized Structural Complexity, $\mathcal{C}_s/\mathcal{C}_s^{*}$, thus effectively `factors out' the sharpness of the first-order-like compositional transition to reveal the dynamics of the structural properties alone. The compositional-normalized form $\mathcal{C}_s/\mathcal{C}_s^*$ is not recommended as a metric for absolute complexity on its own as it represents the degree of structural complexity present as a fraction of what is possible for a given composition (thus, for instance, $\mathcal{C}_s/\mathcal{C}_s^* \rightarrow 1$ for a uniform field). However, around the critical point it will serve as a descriptive metric what effectively isolates the signal we are interested in.

\begin{figure}[t]
    \centering
    % This is the figure generated by the provided Python script
    \includegraphics[width=\columnwidth]{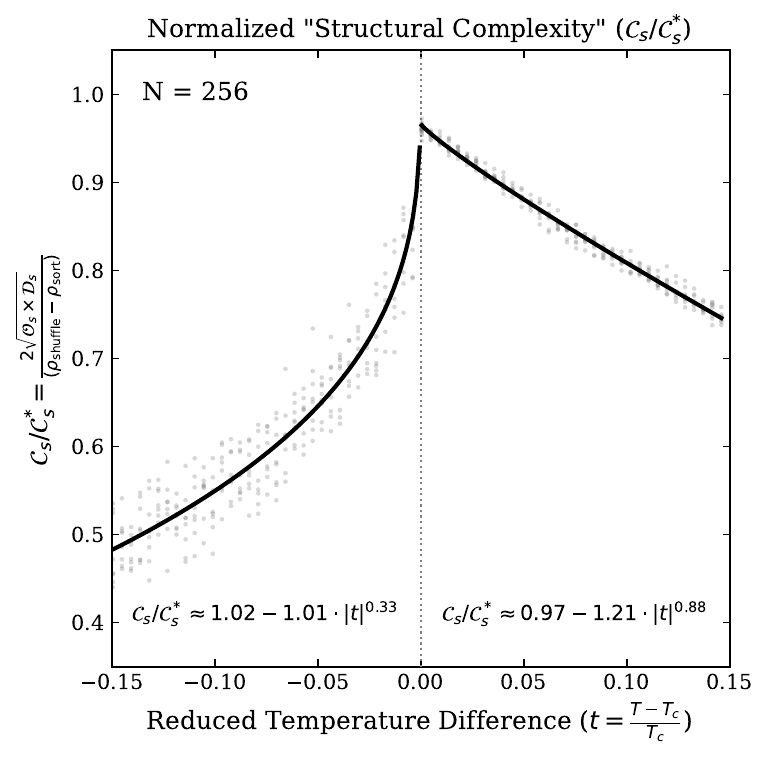}
    \caption{The normalized structural complexity, $\mathcal{C}_s/\mathcal{C}_s^*$, as a function of the reduced temperature, $t=(T-T_c)/T_c$, for a $256 \times 256$ lattice. The data (gray points) are approximated by the analytic form $\mathcal{C}_s/\mathcal{C}_s^* \approx A - B |t|^{\gamma}$ (black lines), with a unique set of parameters for each side of the $T_c$.}
    \label{fig:cs_fit_plot}
\end{figure}

We plot this compositional-normalized metric $\mathcal{C}_s/\mathcal{C}_s^*$ against the reduced temperature $t=(T-T_c)/T_c$ in Fig.~\ref{fig:cs_fit_plot}. We find that the normalized form exhibits a sharp, well-defined peak precisely at the critical temperature. This confirms that our measure of sophisticated structure is maximized at the phase transition. The resulting curve is best described as a ``bent cusp," an asymmetric peak that is finite at its maximum and which seems to approaches a maximum value at the critical temperature asymmetrically from the high and low-temperature sides.

To quantitatively characterize the nature of this peak, we fit the data to a phenomenological scaling function. A simple power law of the form $|t|^{-\gamma}$, which describes the divergence of quantities like magnetic susceptibility ($\chi$) or specific heat ($C_v$) in the thermodynamic limit, is inappropriate here. Such a function tends to $\infty$ at $t=0$, whereas our metric is designed to be non-divergent and our data clearly show a finite peak. To model this, we employ a phenomenological function that captures both the finite peak and the power-law-like decay:
\begin{equation}
    f(t) = A - B |t|^{\gamma}.
    \label{eq:fit_func}
\end{equation}
This function avoids the divergence at $t=0$ through the regularization term `+A', which captures the formation of the characteristic peak and its amplitude. The parameter $\gamma$ serves as an effective critical exponent, modulated by a factor $B$, describing the precipitous decay of the complexity away from the critical point.

While this is a simplified model, it allows for a robust, quantitative characterization of the peak's shape.
We perform separate fits of Eq.~\ref{eq:fit_func} to the data on the low-temperature ($t<0$) and high-temperature ($t>0$) sides of the transition separately. As shown in Fig.~\ref{fig:cs_fit_plot}, this model provides a useful description of the data. The resulting exponents, $\gamma$, quantify the decay rates on either side of the peak while the amplitudes, $A$, describe the maximum value reached at the peak. We note that the result that the amplitudes are very nearly $A \simeq 1$ suggests that the degree of ``structural complexity'' present in the critical Ising state is nearly the maximum possible.

Another key feature of the data, captured by the fit, is the distinct asymmetry of the peak. Even after normalizing away the asymmetry due to the spontaneous magnetization, the high-T side of the cusp is consistently higher and decays more slowly than the low-T side. This asymmetry is not an artifact, but a direct consequence of the multiplicative nature of our metric, $\mathcal{C}_s = \sqrt{\mathcal{O}_s \cdot \mathcal{D}_s}$. The `S-curve' shape of $\mathcal{D}_s$ naturally lifts the high-T side of the peak relative to the low-T side, reflecting the greater degree of entropy present there and resulting in the observed asymmetry. The structural complexity just below $T_c$ arises primarily from the boundaries of small islands, while just above $T_c$ the system is a symmetric collection of interpenetrating domains of both spin types, with a nearly scale-free distribution of sizes. Our second-order complexity metric, $C_s$, is sensitive to this fundamental difference in morphology.

\newpage

\section{Discussion and Future Outlook}
\label{sec:discussion}

In this work, we have introduced and validated a practical, information-theoretic framework for quantifying structural complexity in systems given spatial, ``image-structured'' data, using the 2D Ising model as a canonical testbed. Our central finding is that a normalized, compression-based metric of structural complexity, $C_s$, exhibits a sharp and unambiguous peak precisely at the system's critical temperature, $T_c$, as expected according to the system's known behavior. The success of this ahistorical, model-agnostic method in identifying this key feature of a classic physical system suggests its broad potential as a tool for automated discovery in data-intensive science. In the following section, we discuss the significance of this result, motivate its relationship to other established concepts in statistical mechanics, information theory, and complexity science, and discuss some potential applications of this metric in other scientific domains.

%By extending this analysis across various length scales via the established block-spin coarse-graining procedure, we visualized the phase transition of the 2D Ising model as a ``structural complexity spectrogram." This spectrogram reveals a dramatic, vertical plume of high complexity density that is uniquely present at $T_c$ and spans all measured scales, providing a clear, quantitative fingerprint of the scale-invariance inherent to critical phenomena and clearly differentiating it from the decaying structural complexity near the critical point. 

\subsection{Context within Statistical Mechanics}
The numerical results presented here provide quantitative validation for a concept that is well-established in statistical physics: that a system's internal configuration is maximally ``complex" at its critical point, where the frustration between its competing tendencies toward order and disorder is greatest. Our complexity metric, $\mathcal{C}_s = \sqrt{\mathcal{O}_s \cdot \mathcal{D}_s}$, is constructed to identify this unique state, and can be understood as being composed of empirical proxies for fundamental quantities in statistical physics.
The ``complexity" of the critical state is not a measure of disorder in the traditional thermodynamic sense, but refers to the presence of rich, hierarchical structure which is known to emerge around  criticality.
%At the critical point, the system is scale-invariant; there are correlated domains of spins at all possible length scales, from the lattice spacing up to the size of the system itself \cite{Kadanoff1966}. Away from criticality, a characteristic length scale exists: in the low-temperature phase, it is the infinite size of the single percolating domain; in the high-temperature phase, it is the small size of transient thermal fluctuations. The critical point is unique in its lack of a characteristic scale. Our primary finding is that this physical property of scale-invariance has a direct and measurable information-theoretic signature.

%The first component, Structural Order ($\mathcal{O}_s$), quantifies the system's departure from statistical randomness. As we have noted, its behavior—vanishing at the extremes of temperature and peaking at $T_c$—is qualitatively identical to that of the **multi-information** \cite{Erb2004}. Multi-information measures the total statistical redundancy among all spins, or the "information cost" of the correlations. It is a fundamental measure of how much the system as a whole differs from a collection of independent parts. The fact that our empirically-derived $\mathcal{O}_s$ tracks this formal measure provides a strong validation of our compression-based approach as a probe of system-wide correlations.

However, the second component, Structural Disorder ($\mathcal{D}_s$), which quantifies the system's departure from trivial, crystalline simplicity, serves as a meaningful microstate-level surrogate for the thermodynamic entropy $S(T)$. Both are monotonically increasing functions of temperature with a `S-curve' form that increases most steeply at the critical point. This behavior is 
expected as both quantities are sensitive to the system's departure from order; $\mathcal{D}_s$ in terms of the lack of redundant information in the structure of the microstate and $S(T)$ in terms of the number of microstates with indistinguishable macroscopic properties. 

%The final metric $\mathcal{C}_s$ is thus the geometric mean of a multi-information-like term and a configurational-entropy-like term. It is large only for states that are simultaneously highly correlated (high $\mathcal{O}_s$) and structurally intricate (high $\mathcal{D}_s$). This provides a clear, quantitative criterion for the emergent complexity that is the hallmark of criticality. While traditional observables like magnetization or entropy either vanish or pass through an inflection point at $T_c$, our $\mathcal{C}_s$ provides a direct, peak-based indicator of the transition itself. This reinforces the idea that the critical point is not merely a boundary, but a distinct regime of behavior characterized by a maximal capacity for sophisticated structure.

Furthermore, our observation that the metric $\mathcal C_s$ not only peaks but is also asymmetric at the critical point is consistent with the behavior of other known thermodynamic parameters.
For instance, the amplitudes of the specific heat and magnetic susceptibility are not, in general, equal on both sides of the transition \cite{Aharony1976}. The ``bent cusp" in our $\mathcal C_s$ plot above (Fig.~\ref{fig:cs_fit_plot}) can thus be understood as an information-theoretic manifestation of this same fundamental discontinuity between a symmetry-broken state and a symmetric one.

\newpage
\subsection{Context within Information Theory}

The effectiveness of our approach stems from the impressive ability of lossless compression algorithms to isolate structural information or ``hidden order'' in image data. Our work builds most directly upon the precedent of using Lempel-Ziv (LZ) compression to analyze physical systems \cite{Ziv1977}. Our definition stems from the insight that the size of an optimally compressed file is a practical, empirical estimate of the Shannon entropy or information content of a given data source \cite{Cover2006, Kaspar1987, Aaronson2014, Estevez2015}. 

Our ``Structural Order'' metric ($\mathcal{O}_s$) is defined as the normalized difference between the compressibility of a shuffled lattice and the true lattice: $\mathcal{O}_s =(\rho_{\text{shuffle}} - \rho)$. This quantity is a direct, algorithmic analog of a core concept in information theory: the \textbf{multi-information}, or excess entropy, $I(\sigma)$ \cite{Watanabe1960, Feldman2003}. Multi-information is formally defined as the information-theoretic distance between the full joint probability distribution of the system and the product of its independent marginal distributions:
$I(\sigma) = \left( \sum_{i} H(s_i) \right) - H(\sigma)$,
where $H(s_i)$ is the Shannon entropy of a single spin and $H(\sigma)$ is the joint entropy of the entire system. It quantifies the total amount of statistical redundancy or ``shared information" among all components, vanishing only if all are independent.

Our $\mathcal{O}_s$ metric inherits directly from this definition: the compressibility of the shuffled lattice, $\rho_{\text{shuffle}}$, which has all spatial correlations destroyed, serves as our empirical proxy for $\sum H(s_i)$ while the compressibility of the orinal lattice, $\rho$, is our proxy for the information content of the correlated joint system, $H(\sigma)$. 

The applicability of multi-information to the study of phase transitions is well-established. Foundational work by Arnold \cite{Arnold1996} and later, more rigorously by Erb and Ay \cite{Erb2004}, demonstrated that the multi-information per site in the 2D Ising model behaves as a ``measure of the middle." It vanishes at both zero and infinite temperature and exhibits a sharp peak precisely at the critical temperature, similar to our $\mathcal{O}_s$. However, while the behavior of formal information-theoretic quantities like multi-information is analytically solvable for systems like the 2D Ising model, these methods are fundamentally limited to the small class of systems for which such solutions exist. 

Furthermore, while multi-information (and by analogy, our $\mathcal{O}_s$) is an excellent ``correlation detector," it is an imperfect ``complexity detector." A particularly relevant study by Melchert and Hartmann applied LZ compressability to the 1D \textit{time-series} of a single spin in the Ising model and also found a similarly shaped maximum in their normalized complexity metric at $T_c$ (Fig. 1 in \cite{Melchert2015}). However, as we have shown, such a metric assigns a high value not only to the emergent fractal structure of the critical state but also to simple periodic structures. This is the motivation for introducing our combined metric, $\mathcal{C}_s = \sqrt{\mathcal{O}_s \cdot \mathcal{D}_s}$, which we hope can serve as a detector of complex structure more broadly.

\newpage
\subsection{Context within Complexity Science}

We contrast our proposed data-driven, compression-based approach with other existing formal measures of complexity. While theoretically profound, metrics like Gell-Mann's effective complexity \cite{GellMann1996}, Lloyd's thermodynamic depth \cite{Lloyd1988}, and Crutchfield's statistical complexity ($C_\mu$) \cite{Crutchfield1989} present significant practical barriers for application to static, high-dimensional, statistically heterogeneous data. They respectively require an a priori model of the system's regularities, knowledge of its entire causal history, or a sufficiently repetitive data stream from which to infer a predictive model. Our method deliberately sidesteps these requirements, operating directly on a single spatial snapshot without assumptions. It is not intended to measure the same deeper properties of causality or generative process, but instead provides an operational, intuitively descriptive measure of ``patternedness" that, as we have shown, is a powerful empirical correlate of physical criticality. 

The structural complexity metric proposed in this work, $\mathcal{C}_{s}$, inherits from a storied lineage of theoretical and statistical complexity measures designed to meet the intuitive criterion of vanishing in the limits of both perfect order and perfect disorder. Most notably, the conceptual architecture of $\mathcal{C}_{s}$ is structurally similar to the López-Ruiz–Mancini–Calbet (LMC) complexity \cite{LMC1995}, introduced in 1995. The LMC measure defines complexity ($C_{LMC}$) as the product of two competing quantities: information content (typically Shannon entropy, $H$) and ``disequilibrium" ($D$), which measures the distance of the system's element-wise probability distribution from equipartition. 

Our proposed metric can be understood as a compression-based, spatially aware cousin of the LMC framework, insofar as we choose to meet the naive ``one-hump” criterion by structuring our metric as a product of two competing terms. More specifically, our ``Structural Order" term ($\mathcal{O}_{s}$), which measures the compressibility relative to a maximally randomized (shuffled) state, functions as a loose proxy for the disequilibrium term $D$, capturing the the overall pixel-wise probability distribution's departure from pure randomness (By the LMC definition, ``equilibrium'' here is maximal disorder). Similarly, our counter-term for ``Structural Disorder" ($\mathcal{D}_{s}$), which quantifies the compressibility relative to a perfectly sorted state, serves an intuitively similar purpose to that of the Shannon entropy term $H$ in $C_{LMC}$.

\newpage

While the intuitive inspiration for our metric is reminiscent of $C_{LMC}$, we point out several important functional differences. Firstly, as noted by Feldman and Crutchfield \cite{FELDMAN1998}, $C_{LMC}$ operates only on the level of the probability distribution of the pixel values (ie. the histogram of states, or the magnetization in the case of the Ising model) and thus cannot distinguish between the critical and disordered states which share an equally mixed distribution but are structurally distinct. Our metric, $\mathcal{C}_{s}$, is concerned specifically with the information density extractable from redundant structure and is therefore very sensitive to the difference between a complex critical state and a symmetrically disordered one.\footnote{Said another way: $C_{LMC}$ is invariant under shuffling of the input, whereas our metric $\mathcal{C}_{s}$ is \textit{deliberately} sensitive to the structural difference between an input and its shuffled counterpart. In the language of compressibility, a more fitting analog for $C_{LMC}$ might be $C_{LMC} = H \cdot D \sim (\rho_{\text{shuffle}}) \cdot (\rho^{\text{max}} - \rho_{\text{shuffle}} )$} Therefore, $\mathcal{C}_{s}$ implicitly captures aspects of Effective Measure Complexity \cite{GellMann1996} or Excess Entropy \cite{Feldman2003} (quantities that measure the cost of synchronization or the amount of ``apparent" structure) while remaining computationally tractable for high-dimensional inputs where formal calculation of excess entropy is impossible. Furthermore, our definition of $\mathcal{C}_{s}$ as the geometric mean of two compressibility ratios, $\sqrt{\mathcal{O}_s \cdot \mathcal{D}_s}$, ensures that the metric shares the dimensionality of computable information density, which is intensive with respect to system size (ie. even as $N \rightarrow \infty$, $\mathcal C_s(T_c) \approx 0.5$). We suggest that these properties make $\mathcal C_s$ a desirable, well-behaved, complexity measure, and we emphasize that its ability to simultaneously identify the critical peak in this canonical example and to distinguish between `crystalline' and `fractal' order more broadly suggests that it faithfully reflects a meaningful property of the input structure.  

%Conversely, our ``Structural Order" ($\mathcal{O}_{s}$), which measures the compressibility relative to a maximally randomized (shuffled) state, functions as a proxy for the disequilibrium term $D$, capturing the departure from randomness. While LMC complexity typically utilizes the simple product $H \times D$, we employ the geometric mean $\sqrt{\mathcal{O}_s \cdot \mathcal{D}_s}$ to ensure the metric shares the dimensionality of a compressibility ratio, but the phenomenological behavior remains identical: the metric is convex with respect to randomness, peaking only in the intermediate regime where both structure and information content are non-trivial. 

The science of complexity is still in its adolescence and is wanting for new tools and frameworks for exploration. Thus, we propose $\mathcal{C}_{s}$ not as a replacement for rigorous measures like Statistical Complexity ($C_\mu$), which quantify the state-space size of the underlying causal generator, but as a pragmatic, finite-time computable heuristic. We hope it can help bridge the gap between abstract complexity theory and experimental application, providing a loose but instantly-applicable gauge for criticality that requires no knowledge of the system's Hamiltonian or order parameters.  While we do not expect that our phenomenological description will be a fruitful baseline for much future theoretical work, we hope that the practical efficiency of our metric will enable data-driven research and discovery of complex phenomena in other frontier scientific domains.

\newpage

\subsection{Future Outlook}
The validation of our compression-based structural complexity metric, $\mathcal{C}_s$, as a robust indicator of the critical phase transition opens several promising avenues for future research. While our current definition successfully resolves the ambiguity between simple periodic order and the ``emergent'' complexity of the critical state, further refinements could yield an even more discerning probe. 

%As we have discussed, in the limit of $N \rightarrow \infty$, our metic resembles a geometric mean of the element wise mutual information and thermodynamic entropy. However, for finite size data, our metric incurs some imprecision due to the nature of the compressed files. Compression algorithms like PNG include a ``header'' to instruct client software on how the image data is to be decoded. For very large files, this header volume is negligible, yet it can constitute an impactful percentage of the compressed size of smaller images. We account for this approximately in our method by subtracting $\rho_{\text{sort}}$ to get the disorder term. It may however be fruitful to explore alternate definitions of the ``baseline of compressibility'' that might bring our metric more in line with established theoretical properties.

\vspace{12pt}

One compelling direction is to explore other, perhaps more context-motivated compression methods, such as Neural Network embeddings (ie. autoencoders, see \cite{Hinton2006}), singular value decompositions, or other more ``lossy'' compression algorithms (e.g. JPEG). It could also be interesting to exchange the notion of upper-bounding the true Kolmogorov complexity for a more flexible notion of a time-bounded or data-limited Kolmogorov complexity, approximating how a finite observer (like a Human or Artificial Neural Net) would experience a given system.

\vspace{12pt}

Perhaps the most exciting extension of this framework is to move beyond a single complexity value and analyze the system's structure as a more descriptive function of locality or scale:

Due to its construction as a property of the entire image, $\mathcal C_s$ is blind to \textit{where} the complex structure is localized within image. It is also possible for the metric to be confused by images that contain distinctly separate regions of order and randomness.  In real data, ``complex'' structures may occupy only certain areas of the input field (for instance: raw tissue samples or galaxy cluster images) and my contain ordered and disordered areas simultaneously. We suggest that, for automated application to large real-word datasets, it may be useful to implement a `scanning' procedure which sequentially considers only a small window of the image. We explore this challenge in more detail in Appendix A.

Furthermore, it may also be interesting to explore an extension of $\mathcal C_s$ which is sensitive to some notion of physical scale. The critical state is defined by its scale-invariance, and a meaningful measure of critical complexity should reflect this property. By combining our metric with a coarse-graining procedure inspired by Kadanoff's block-spin renormalization group \cite{Kadanoff1966}, one can construct a ``structural complexity spectrum," $\mathcal C_s(L)$, which quantifies complexity at different present at and above a given length scale $L$. In such a spectrum, the scale-invariant nature of the critical point manifests as a high, broad plateau, distinguishing it from non-critical states whose complexity is either near-zero or confined to a narrow range of scales. We explore a preliminary version of this spectral analysis some of and present some of our initial findings in Appendix B. 

\newpage 

Beyond refining the metric itself, there are several promising avenues for applying this framework. The universality principle suggests that this method should be equally effective at identifying the critical points of other models in the same universality class, such as the lattice gas or binary alloy models. Applying it to other simulated or theoretical systems in different universality classes would be an interesting test of its generality. The most exciting applications, however, lie in the analysis of real-world experimental data. As discussed in the introduction, this structural complexity metric could be applied to datasets from domains as diverse as neuroscience, geology, and astrophysics as a powerful, model-agnostic tool for automated feature detection and the discovery of critical behavior in systems where the underlying source is unknown or impossible to model.

The work presented here on a canonical physical model serves as a benchmark validation for these future explorations. The methodology is directly applicable to any dataset that can be represented as a family of 2D images. The potential applications are vast, to provide a few examples: In materials science, $\mathcal C_s$ could provide an automated method for quantifying the microstructural complexity of alloys or composites from electron micrographs. In medicine, it could be used to analyze the complexity of cortical folding patterns or automatically detect the boundary of heterogeneous tissue which are often indicative of impending cancer \cite{Irshad2014}. And in astrophysics, it could help automatically classify galaxy morphologies, analyze the different regimes of turbulence in interstellar gas clouds, or quantify the complex nature of the filamentary structure of the cosmic web. The strength of this approach lies in its input universality and computational efficiency. As scientific disciplines become increasingly data-rich, the need for robust, automated, and model-agnostic tools to discover and quantify emergent structure will only grow. This work provides a validated, physically-grounded step in that direction.

\section*{Acknowledgments} \label{sec:Acknowledgments}
%The authors are grateful for conversation and guidance from [---------------]. 
The authors are grateful to James Crutchfield, Stephen Shenker, and Omar Aguilar for helpful comments.

This work utilized computational resources from the SLAC National Accelerator Laboratory's dedicated high-performance compute cluster, S3DF.  

Meaningful software development and a preliminary literature review was performed using a Large Language Model, Gemini 2.5, made available to the authors through Stanford University's pilot ``AI Playground." All LLM-generated code was verified and extensively tested by the authors. No LLM tools were used to interpret results or write text for this document.

C.J.~gratefully acknowledges fellowship support from the National Science Foundation's Graduate Research Fellowship Program (GRFP) and Stanford University's William R. Hewlett Graduate Fellowship.

\newpage

\newpage
\bibliography{apssamp}% Produces the bibliography via BibTeX.

\newpage

\section*{Appendix A: Locality Blindness}

The structural complexity metric, $\mathcal{C}_s$, as defined and utilized in the main body of this work, is a global quantity. As a consequence of the construction of $\mathcal C_s$ in terms of the compressibility of the features of the \textit{entire} image relative to its size, the metric is ``locality blind.'' Both of the component factors, Structural Order ($\mathcal{O}_s$) and Structural Disorder ($\mathcal{D}_s$), are insensitive to the spatial position of features within the image, so the product $\mathcal C_s$ can be confused by images with spatial inhomogeneity. This global averaging can lead to ambiguous or misleading results for systems composed of distinct regions with qualitatively different types of order.

\begin{figure}[h] % [t!] is a placement hint: "try top, but be insistent!"
    
    \centering % Center the entire stack of images horizontally

    \includegraphics[width=0.95\linewidth]{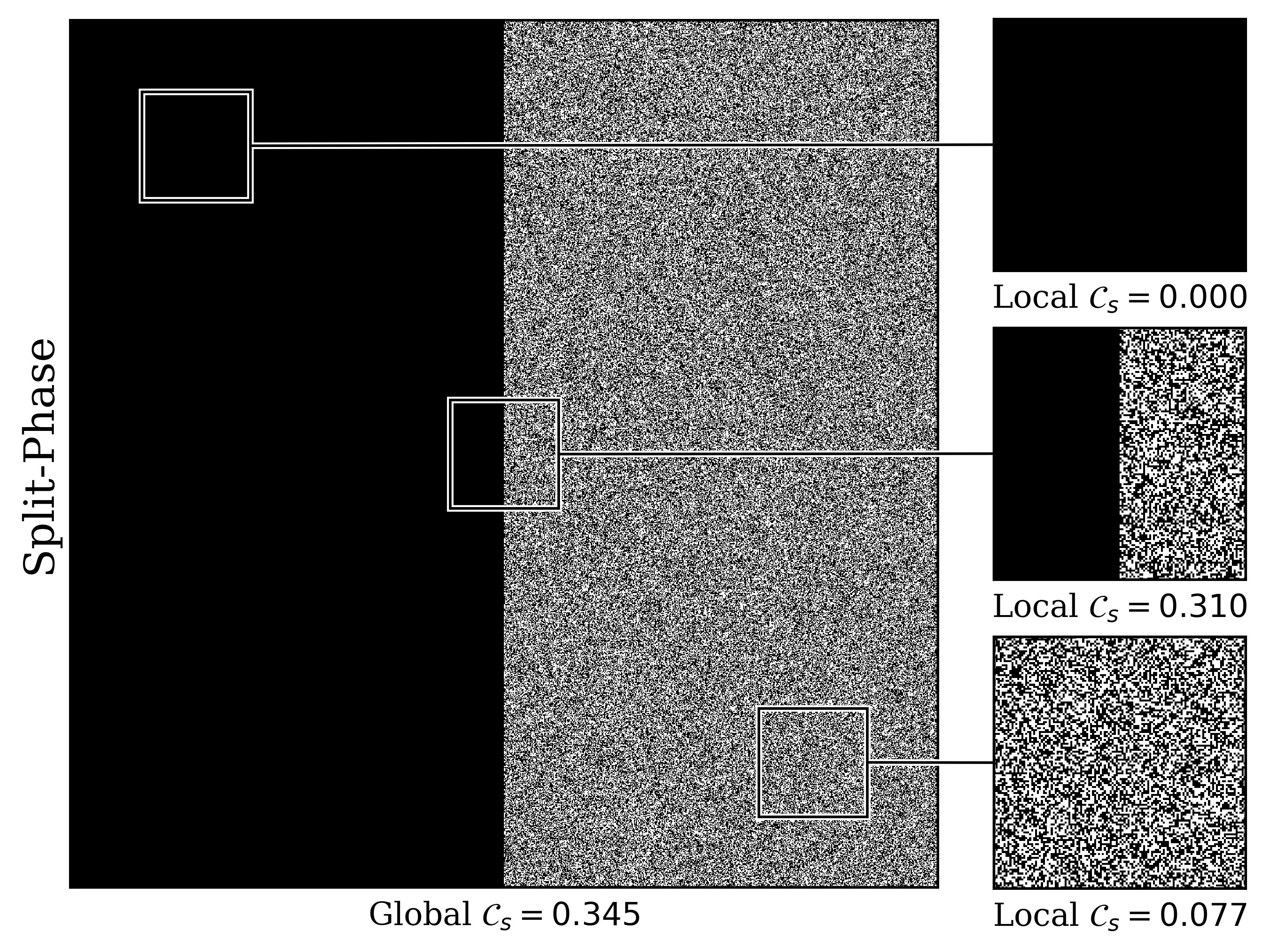}
    \includegraphics[width=0.95\linewidth]{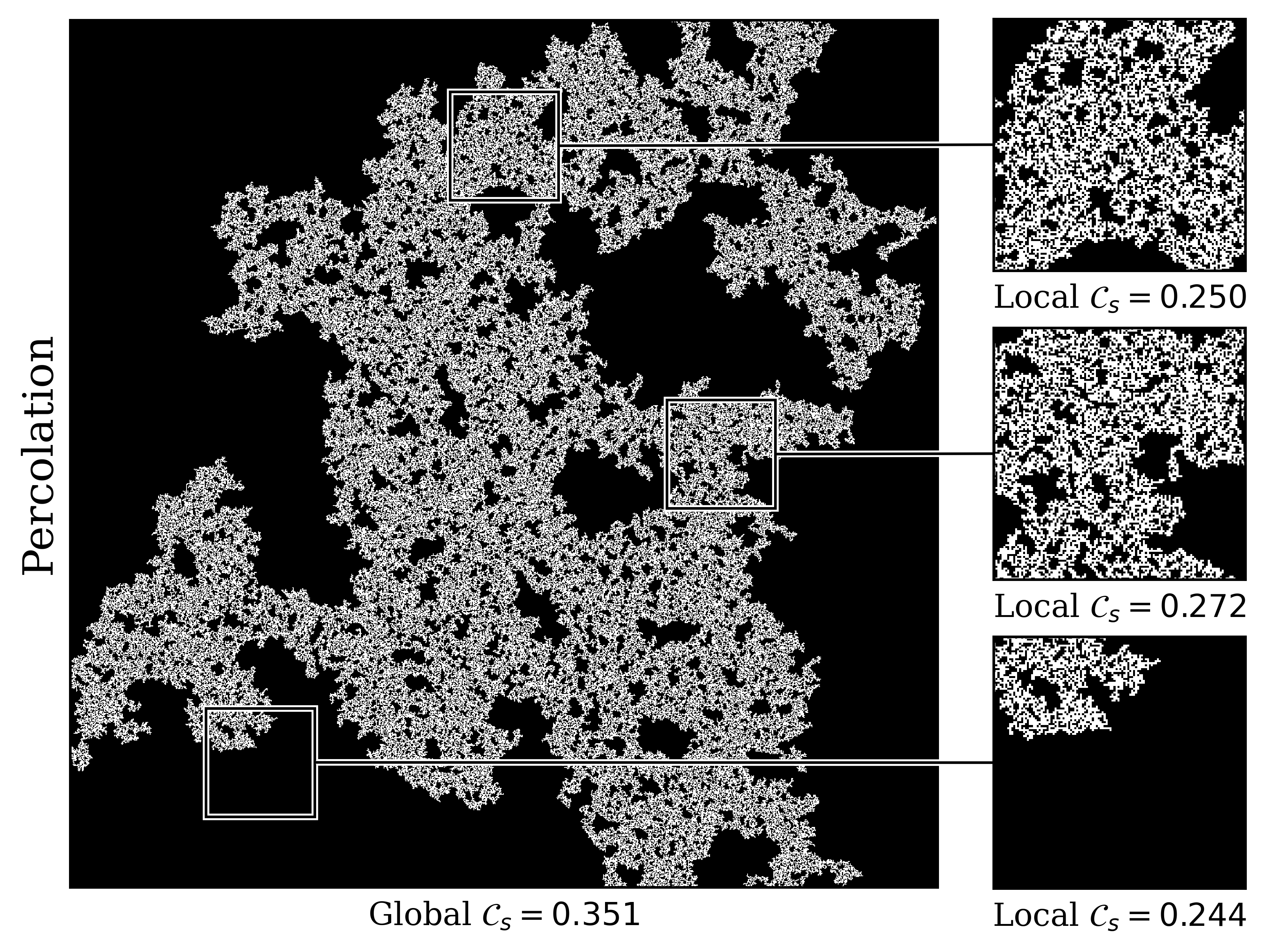}
    \includegraphics[width=0.95\linewidth]{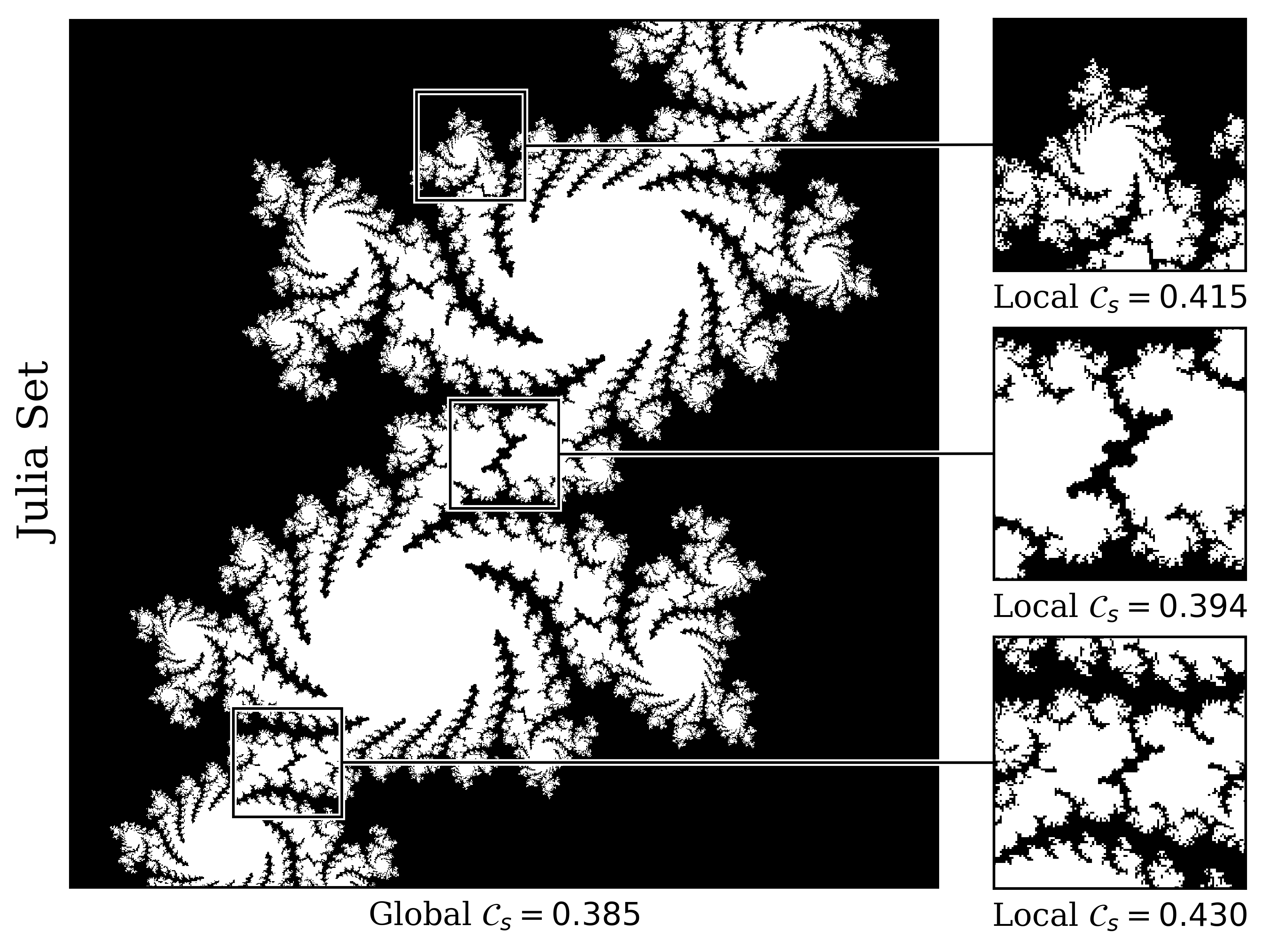}

    % --- A single caption and label for the entire figure stack ---
    \caption{
        Example measurements of $\mathcal C_s$ for some notable configurations and their subset zooms. The global value of $\mathcal C_s$ for the entire image and the values of $\mathcal C_s$ for for the local ``chunks'' are shown below their respective snapshots. 
        %The values of $\mathcal C_s$ for for the local ``chunks'' are shown beneath a snapshot zoom in a column to the right of the main image. 
        %Examples include a discretized realization of the Julia set, a so-called ``Percolation'' fractal which is grown stochastically from a seed cell, and a ``split phase'' lattice with a domain wall separating a perfectly ordered phase from a maximally disordered one. 
    }
    \label{fig:local_zoom}

\end{figure}

To illustrate this limitation, we consider an artificial ``split-phase" configuration where one half is perfectly ordered and the other half is maximally random. The global $\mathcal{C}_s$ value for this image is moderate, failing to capture the fact that it is composed of two regions which, individually, have near-zero complexity. The metric averages the simple order of one half with the simple disorder of the other, producing a single number that does not accurately describe either. Figure~\ref{fig:local_zoom} demonstrates this effect. While the global $\mathcal{C}_s$ for the split-phase lattice is non-zero, an analysis of local ``chunks" reveals the true picture: a chunk taken from the ordered region has $\mathcal{C}_s \approx 0$, a chunk from the random region has $\mathcal{C}_s \approx 0$, and only a chunk spanning the boundary between them exhibits any significant complexity. We contrast this with two example systems containing regions of perfect order with fractal boundaries.

This issue is not limited to artificial cases; real-world systems, from geological formations with distinct strata to brain tissue with both healthy and dysplastic regions, are often inhomogeneous.
For applications that require a more discerning, spatially-resolved summary of complexity, a different approach is needed. We propose a ``convolutional" pipeline inspired by the architecture of convolutional neural networks (CNNs) used in computer vision. We suggest that it may be useful to  transform our global $\mathcal C_s$ metric into a local one by applying it to a moving window that scans across the image. The image can be broken up into (overlapping) smaller chunks and fed into the $\mathcal C_s$ compression calculator individually, thereby yield a map of $\mathcal C_s$ as a function of chunk position. 

This convolutional approach introduces a new free parameter: the window size, $W$. We suggest that this parameter will likely need to be tuned for specific applications to meet the needs of the scientific problem: A larger chunk size will be less dominated by finite size effects and have greater opportunity to extract recursive structure, while a smaller chunk size will be more localized and generate a more fine-grained map of the distribution of complex structure within the image at the cost of being more sensitive to noise and artifacts. 
The optimal choice of $W$ will therefore be application-specific, depending on the characteristic length scales of the features one wishes for an experiment to be sensitive to.

\newpage

{\color{white} white space}

%As demonstrated in Fig.~\ref{fig:local_zoom}, this local complexity mapping correctly identifies the boundary of the split-phase lattice as the sole region of interest. For the Julia set and percolation fractals, it reveals that complexity is not uniformly distributed but is concentrated along the intricate boundaries, providing a much richer characterization. 
%This ``convolutional'' framework represents a powerful future direction, transforming our scalar metric into a full-fledged image analysis tool capable of automatically segmenting and identifying regions of emergent structure within complex spatial data.

\begin{figure*}[t]
  \centering
  \includegraphics[width=\textwidth]{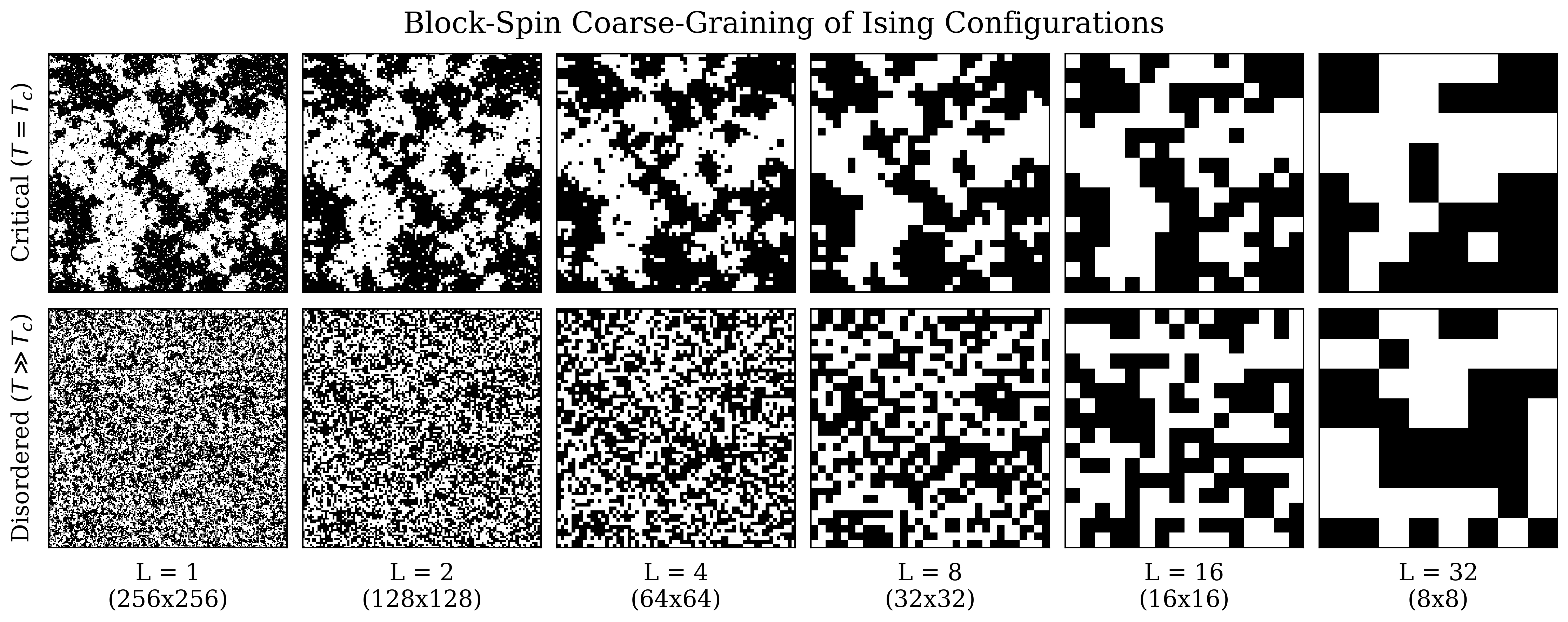}
  \vspace{-12pt}
  \caption{A schematic of the block-spin coarse-graining procedure. The original lattice (left) is partitioned into $L \times L$ blocks (here, $L=3$). The spins within each block are averaged, and a single ``super-spin" is assigned to the new, smaller lattice (right) based on a majority rule. This transformation allows us to view the system's structure at a coarser length scale.}
  \label{fig:block_spins}
\end{figure*}

\newpage

\section*{Appendix B: A Complexity ``Spectrum''}
\label{sec:spectrum_method}

While a single complexity value for a given configuration is informative, the physics of criticality is fundamentally a story of scale. At the critical point, the system is scale-invariant, meaning it ``looks the same" at all levels of magnification. To probe this phenomenon directly, we require a method that can measure structural complexity not as a single number, but as a function of length scale, $L$. This leads us to define a ``structural complexity spectrum," $\mathcal C_s(L)$. We construct this spectrum using a coarse-graining procedure directly inspired by and in accordance with the system's Renormalization Group.

\vspace{-10pt}

\subsubsection{Coarse-Graining and the Renormalization Group}

The Renormalization Group, developed by Kadanoff and formalized by Wilson, is the theoretical framework for understanding phase transitions and critical phenomena \cite{Kadanoff1966, Wilson1974}. The central idea of RG is to systematically ``zoom out" from a system, observing how its effective parameters change with scale. At a critical point, the system is a fixed point of this transformation; its statistical properties are invariant under changes in scale.

Kadanoff's original ``block-spin" thought experiment will provide the direct physical intuition for our method \cite{Kadanoff1966}. The procedure involves partitioning the lattice into blocks of side length $L$ and replacing each block with a single ``super-spin" representing the majority state of the original spins within that block (see Fig.~\ref{fig:block_spins}). This transformation creates a new, smaller lattice whose spins represent the state of the original system at a coarser resolution. By repeating this for a range of block sizes, we can study the system's structure at different scales.

%This technique has long been a cornerstone for understanding the Ising model. It provided the first conceptual path to calculating critical exponents and understanding the universality of critical behavior long before it was used for direct numerical analysis. More recently, coarse-graining techniques, often through real-space RG methods, continue to be a powerful tool for analyzing spin systems, neural networks, and other complex models where scale-invariance is a key feature \cite{Swendsen1979, Mehta2014}.

\subsubsection{Relation to Signal Processing and Image Compression}

The block-spin coarse-graining procedure, while rooted in statistical physics, is conceptually and mathematically related to several standard techniques in signal processing and image compression. 
%Understanding these connections provides a deeper intuition for what our complexity spectrum, $\mathcal C_s(L)$, is measuring.
At its core, the block-spin transformation is a form of \textbf{downsampling and low-pass filtering}. Each $L \times L$ block is replaced by a single value representing the block's average property (in our case, the majority spin). This is functionally similar to first blurring the image to remove details smaller than the scale $L$, and then subsampling the blurred image to reduce its resolution. In this context, the block-spin method can be seen as a physically motivated, structure-preserving flavor of common image compression techniques. Like common blurring or down-sampling, it is a real-space low-pass filter. However, in the spirit of the Renormalization Group, it guarantees that the composition of the resampled images is directly comparable to their original formats.

\begin{figure}[b]
  \centering
  \includegraphics[width=0.95\linewidth]{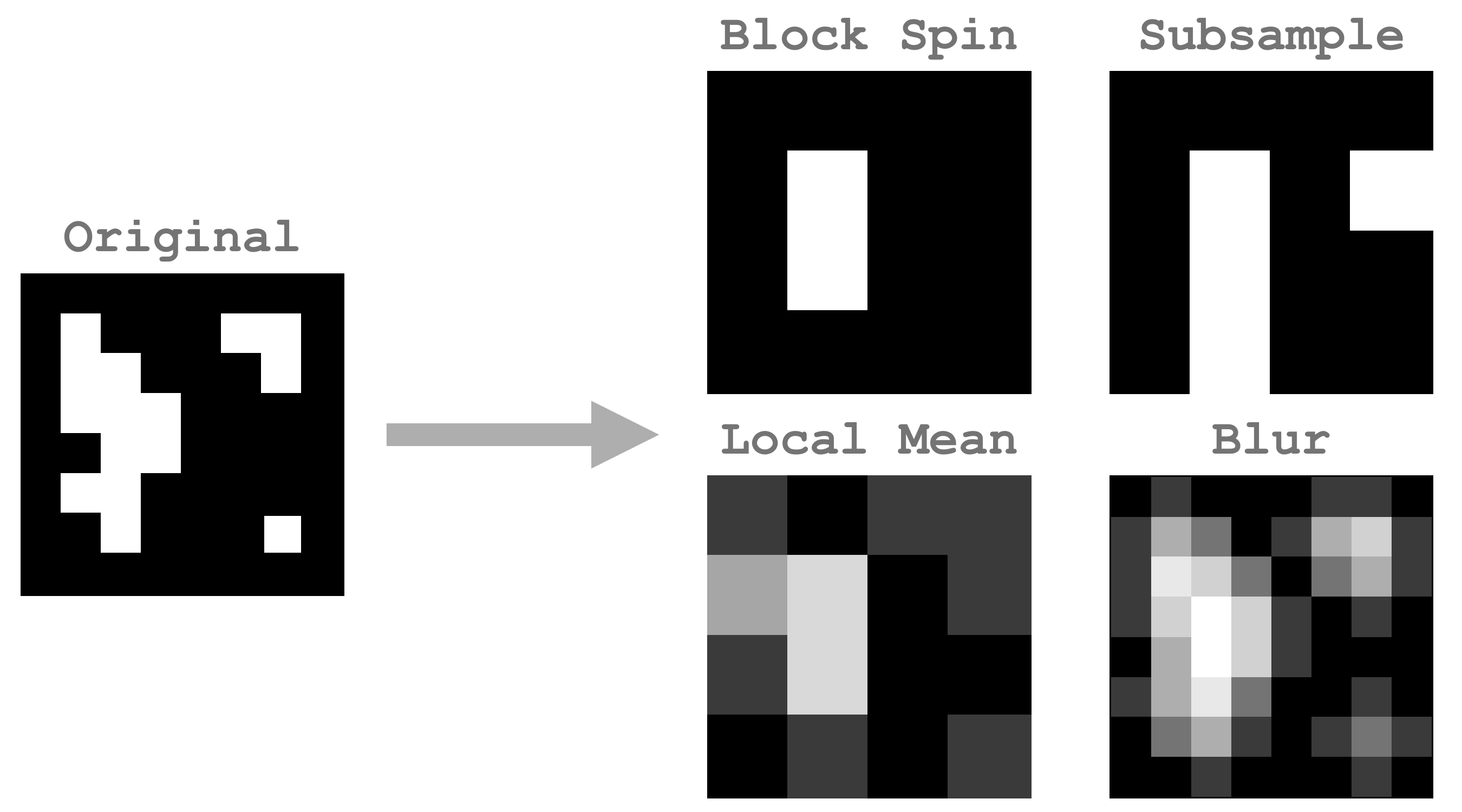}
  \caption{Comparison of example low-pass filters.}
  \label{fig:compress_cartoon}
\end{figure}

\begin{figure}
  \centering
  \includegraphics[width=\linewidth]{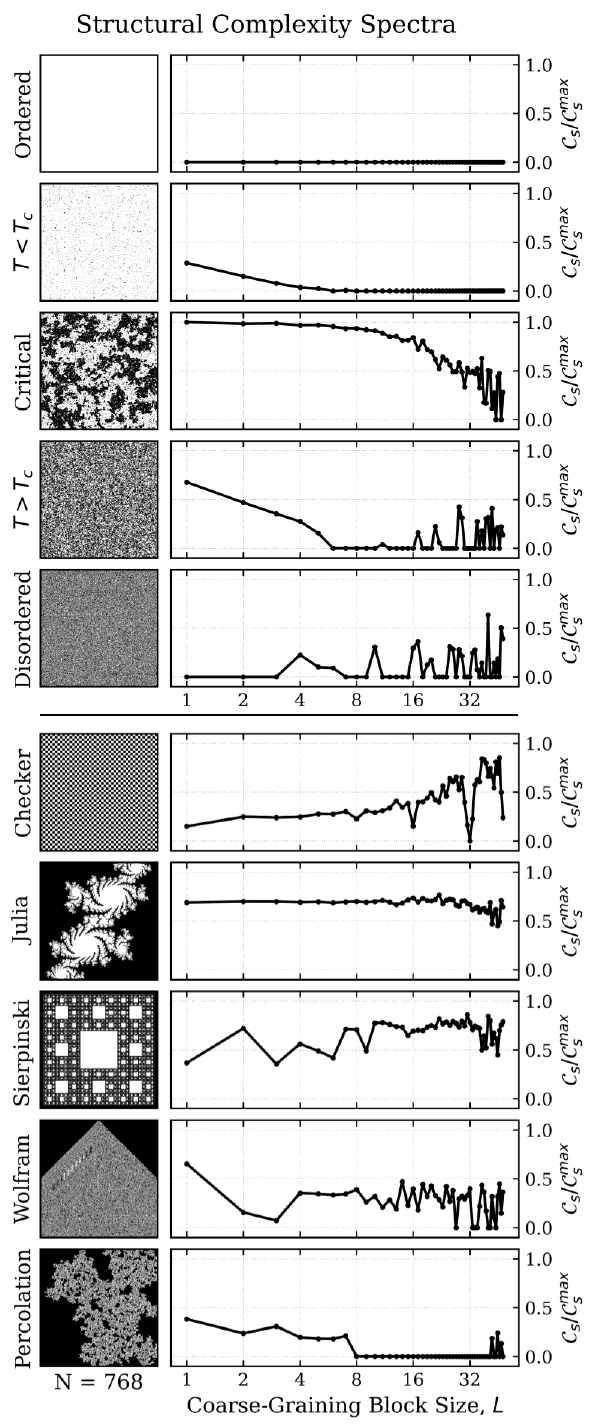}
  \caption{The Block-spin Structural Complexity Spectra $\mathcal C_s(L)$ for some example lattice configurations with $N=256$. The upper five are results of  Ising lattice Markov Chain simulations, while the lower five present other ``out of distribution'' lattice configurations meant to examine the descriptive power of our proposed metric.}
  \label{fig:Unique_Lattice_Spectra}
\end{figure}

\subsubsection{The Block-Spin Complexity Spectrum, $\mathcal C_s(L)$}

We leverage this powerful physical idea to define our complexity spectrum. Following \cite{Martiniani2020}, instead of tracking how the Hamiltonian's coupling constants change with scale, we track how the \textit{structural complexity} changes. Our procedure is as follows. For a given $N \times N$ initial Ising configuration, $\sigma_0$, and a chosen scale, L:
\begin{enumerate}
    \item \textbf{Coarse-Grain:} Partition $\sigma_0$ into blocks of size $L \times L$ and apply a majority rule to generate a new, smaller lattice, $\sigma_L$, of size $(N/L) \times (N/L)$. 
    \item \textbf{Measure Compressibilities ($\rho$):} For this new coarse-grained lattice, $\sigma_L$, we calculate its structural complexity using our compression-based metric. As before, we compute its compressibility, $\rho[\sigma_L]$, and that of its sorted and randomly shuffled counterpart, $\rho_{\text{sort}}[\sigma_L]$ and $\rho_{\text{shuffle}}[\sigma_L]$. 
    \item \textbf{Calculate Complexity $\mathcal C_s(L)$:} The structural complexity at scale $L$ is then simply defined as:
    
\begin{equation*}
    \mathcal C_s(L) = \sqrt{(\rho_{\text{shuffle}}[\sigma_L] - \rho[\sigma_L]) \times (\rho[\sigma_L] - \rho_{\text{sort}}[\sigma_L])}
    \label{eq:cs_L_def}
\end{equation*}
        
\end{enumerate}
By repeating this procedure for a range of block sizes $L$, from small scales (e.g. $L=1$) to much larger scales (e.g. $L=N/4$), we generate the complexity spectrum, $\mathcal C_s(L)$. %The spectra resulting from this procedure for several test configurations are shown here in Figure \ref{fig:Unique_Lattice_Spectra}. 

\vspace{12pt}

To account for (some of) the inevitable finite size effects, we note that the maximum possible value $\mathcal C_s(L)$ can assume is itself a finite function of $N/L$. This value, which we denote $\mathcal C_s^{\text{max}}(L)$, is the one half the difference between the shuffled and sorted compressibilities ($\rho$) of a lattice of size $N/L$:

\begin{equation*}
    \mathcal C_s^{\text{max}}(L) = \frac{1}{2} \Big(  \rho_{\text{shuffle}}^{\text{max}}(N/L) - \rho_{\text{sort}}^{\text{max}}(N/L) \Big)
\end{equation*}

\noindent
Where $\rho_{\text{shuffle}}^{\text{max}}(N/L)$ and $\rho_{\text{sort}}^{\text{max}}(N/L)$ correspond to a lattice of size $N/L$ of equal composition (ie. and equal number of black and white cells). Note that this is not the same quantity as used to normalize Fig~\ref{fig:cs_fit_plot}, which is also a function of the composition of the input. 

We plot the structural complexity spectra normalized by their scale-dependent maximum ($\mathcal C_s / \mathcal C_s^{\text{max}}(L)$) for several Ising configurations and other test cases in Fig.~\ref{fig:Unique_Lattice_Spectra}.
This spectrum has an intuitive physical interpretation. A high value of $\mathcal C_s(L)$ indicates that, even at a resolution of $L$, the system still possesses significant non-random spatial structure.
The shape of the $\mathcal C_s(L)$ spectrum thus serves as a direct, quantitative signature of the system's underlying scaling properties, providing an informative, information-theoretic window into the structure of physical systems at and around criticality.

\newpage

\subsection*{Behavior of the Complexity Spectrum}

Before applying our analysis to the full temperature sweep of the Ising model, we first validate the behavior of the block-spin complexity spectrum, $\mathcal C_s(L)$, on a set of benchmark configurations with known structural properties. These tests are crucial for building intuition and confirming that the metric correctly identifies and distinguishes different types of order and complexity. The results for several key configurations are shown in Fig.~\ref{fig:Unique_Lattice_Spectra}.

\subsubsection{Ising Model Configurations}

We begin by examining configurations drawn from the equilibrium distribution of the 2D Ising model itself, representing several points in its phase transition:
\begin{itemize}
    \item \textbf{Ordered State ($T \ll T_c$):} For a low-temperature, fully magnetized lattice, the spectrum is trivial. The configuration is a uniform field of spins, and any coarse-grained version of it remains uniform. As both the original and shuffled configurations are maximally simple, $\mathcal C_s(L)$ is zero for all scales $L$.
    \item \textbf{Disordered State ($T \gg T_c$):} For a high-temperature, random lattice, the configuration lacks any spatial correlations. The coarse-graining procedure simply produces a new, smaller random lattice. The structural complexity $\mathcal C_s(L)$ is therefore approximately zero across the entire spectrum. 
    \item \textbf{Slightly Ordered State ($T < T_c$):} For a configuration just below the critical temperature, only very small islands of flipped spins can exist. The spectrum shows a very small degree of complexity at the smallest block sizes ($L \lesssim \xi$), which is almost immediately suppressed to zero for larger $L$.
    \item \textbf{Slightly Disordered State ($T > T_c$):} For a configuration slightly above the critical temperature, the correlation length, $\xi$, is finite and small. The spectrum shows a notable amount of structural complexity at the smallest block sizes, which then eventually decays to zero for larger $L$.
    \item \textbf{Critical State ($T = T_c$):} The spectrum for the critical configuration is qualitatively distinct from all others. It exhibits a high, broad plateau, with a nearly maximal $\mathcal C_s(L)$ value extending across a wide range of length scales. This is a definitive signature of scale-invariance. It quantitatively demonstrates that at criticality, non-trivial, complex structure exists at every level of magnification, from the smallest blocks up to system-spanning sizes. This plateau is the central feature we seek to identify.
\end{itemize}

\subsubsection{Geometric and Fractal Patterns}

To demonstrate the generality of our metric beyond the Ising model, we also test it on a suite of artificially generated lattices with more diverse structures. These ``out-of-distribution" tests probe the metric's response to periodic, chaotic, and fractal order.

First, we consider a simple periodic pattern: a checkerboard composed of large $16 \times 16$ blocks. Its complexity spectrum begins small but non-zero, due to the mixed composition and high degree of regularity, but tends upwards for higher $L$ due to mismatch between the checkerboard size and the block size. The metric correctly identifies the characteristic length scale of the pattern, as harmonic dips in the spectra at $L=8,16,32$. This confirms that $\mathcal C_s(L)$ can serve as a meaningful ``structure" spectrometer, measuring what spatial scales contain reducible structure. 

Next, we analyze several classic examples, which, like the critical Ising model, are expected to possess structure at multiple scales:
\begin{itemize}
    \item \textbf{Sierpinski Carpet:} A famous, deterministic fractal created by recursively dividing a square into nine sub-squares and removing the central one.
    \item \textbf{Julia Set:} A fractal defined in the complex plane, generated by iterating coloring points according to whether or not the function $z \to z^2+c$ diverges. 
    \item \textbf{Wolfram's Rule 30:} A one-dimensional cellular automaton \cite{Wolfram1983} known for producing ``deterministic chaos." We stack its temporal evolution to create a 2D lattice exhibiting a mix of regular, triangular structures and apparent randomness.
    \item \textbf{Percolation:} A stochastic fractal formed by randomly occupying neighboring sites on a grid at a critical probability, starting from a seed and resulting in a single cluster that spans the domain.
\end{itemize}

When we compute the complexity spectra for these fractal test cases, we find hints of a unifying result. Despite their vastly different generative rules and visual appearances, the Sierpinski and Julia configurations produce spectra that are notably similar to the critical Ising state: they exhibit a broad region where $\mathcal C_s(L)$ is \textit{roughly} constant and notably greater than zero. The presence of this extended, non-zero plateau strongly suggests that our metric is successfully identifying the fundamental property of scale-invariance shared by these objects. The spectra for Rule 30 and the percolation fractal are more intricate, reflecting their mixed nature and suggesting that the randomness of these objects may be localized on particular scales. These tests suggest that, with some refinement, our compression-based approach could serve as a general-purpose method for detecting and quantifying the ``fractal-ness" or multi-scale complexity of other spatial datasets.

\subsection*{The Temperature--Scale Landscape}
\label{sec:spectrogram_results}

The one-dimensional complexity spectra, $\mathcal C_s(L)$, reveal the scale-dependent structure of the system at fixed temperatures. To capture the full picture of how structure emerges and dissolves across the entire phase space, we now present our central result as a two-dimensional ``complexity spectrogram\footnote{We use the term ``spectrogram'' here in a broad sense, and we suggest that in future applications such a spectrogram might be generated by plotting scale against, for instance, parameters like Reynolds number, packing density, metallically, etc., or indeed as scale vs. time, as is traditional.}." This plot visualizes our $\mathcal C_s$ metric as a function of both temperature, $T$, and coarse-graining scale, $L$, providing a more informative map of the system's complexity landscape.

\begin{figure}[t]
  \centering
  \includegraphics[width=0.95\linewidth]{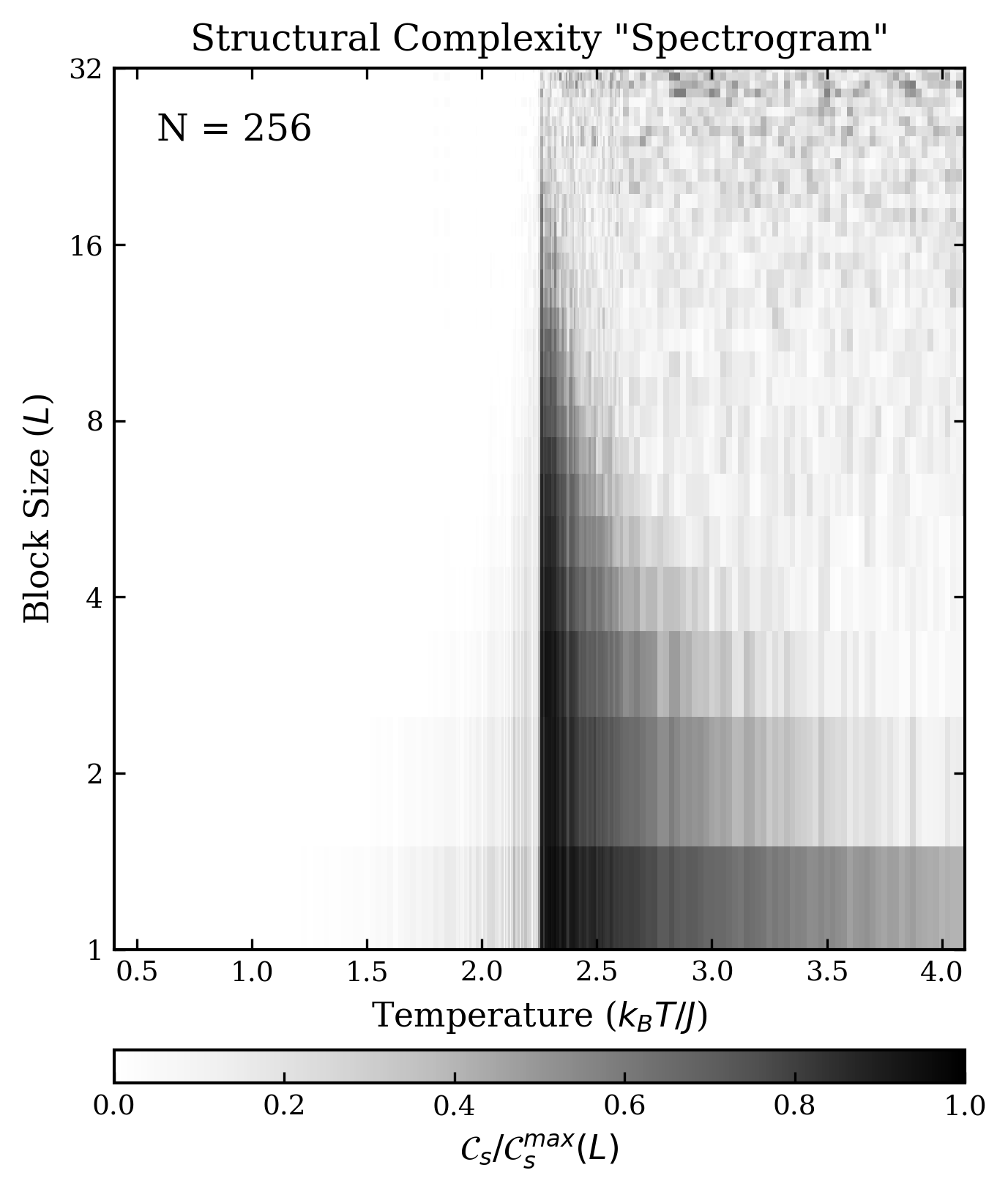}
  \caption{Visualization of the temperature-dependence of the Ising block spectrogram. Cells are colored according to the mean $\mathcal C_s$ value of a family of 16 simulations of temperature $T$, coarse-grained to scale $L$. We increase the temperature resolution from $\Delta T = 0.035$ to $\Delta T = 0.01$ in the range $2.0 < T<2.5$ to better observe the transition.}
  \label{fig:ising_block_spectrogram}
\end{figure}

For this analysis, we utilize the full dataset generated from our high-resolution temperature-sweep of simulations on a $256 \times 256$ lattice. For each of the $16$ independent configurations generated at each temperature point, we compute the block-spin spectrum, $\mathcal C_s(L)$, across a dense set of block sizes, $L \in [1,32]$. The results are then averaged over the 16 runs to obtain a mean value for each metric at every $(T, L)$ coordinate. These mean values are then assembled into a logarithmic 2D grid and plotted as a color map, as shown in Fig.~\ref{fig:ising_block_spectrogram}.

We again display the structural complexity metric, $\mathcal C_s(L,T)$ normalized by the \textit{maximum possible} value of $\mathcal C_s$ at a given coarse-graining scale (based on the shuffled and sorted compressibilities of a \textit{equal-composition} black and white image of the same size). Fig.~\ref{fig:ising_block_spectrogram} shows the mean of the normalized quantity $\mathcal C_s(L,T) / \mathcal C_s^{\text{max}}(L)$ plotted on a dense grid of values $L,T$. The vast majority of the parameter space is empty, or at least noisy and otherwise unstructured, corresponding to $C_s \approx 0$. All of the interesting behavior is concentrated in a narrow region surrounding the critical temperature. 

The shape of this complexity landscape can be intuitively understood by examining its cross-sections. A horizontal slice at the smallest possible block size ($L=1$, the bottom edge of the plot) corresponds to the one-dimensional $\mathcal C_s(T)$ curve discussed previously. It shows a peak of some width maximized at $T_c$. As we move vertically up the plot to larger block sizes $L$, we are effectively viewing this same complexity peak through a progressively coarser lens, and we find that the peak gets asymptotically thinner.

For temperatures away from criticality ($T \neq T_c$), the system has a finite correlation length, $\xi$. Once the block size $L$ becomes larger than $\xi$, the coarse-graining procedure averages away all existing correlations, and the structural complexity $\mathcal C_s(L, T)$ rapidly decays to zero. This results in the ``funnel" shape seen in the spectrogram, where the region of high complexity narrows as $L$ increases.

\newpage

However, along the vertical line at $T=T_c$, the complexity does not decay, or at least not immediately. Instead, we observe a noticeable ``kink" in the funnel's boundary. The value of $\mathcal C_s$ remains nearly maximal and roughly constant for several length scales before it, too, falls off due to the diminished image size. This vertical plume of high complexity a signature of the critical state's established scale-invariance. At criticality, the correlation length is infinite, and thus there is meaningful, non-random structure at every length scale. The block-spinning transformation of a critical lattice configuration produces a new lattice that is itself critical, preserving its high structural complexity.

The eventual decay of $C_s(L, T_c)$ at very large $L$ is an inevitable consequence of finite-size effects. As we coarse-grain the lattice, the resulting system becomes progressively smaller. For example, on a $256 \times 256$ lattice, a block size of $L=16$ results in a tiny $16 \times 16$ system. As established in our finite-size scaling analysis discussion (Fig.~\ref{fig:N_comp}), such small systems cannot support a sharply defined critical peak. Therefore, the decay of complexity along the vertical axis of the spectrogram at $T=T_c$ can be interpreted as a manifestation of finite-size effects: as our ``renormalized" system becomes smaller, the structural signal it can support eventually becomes weaker.

\newpage

\begin{figure*}
  \centering
  \includegraphics[width=0.75\textwidth]{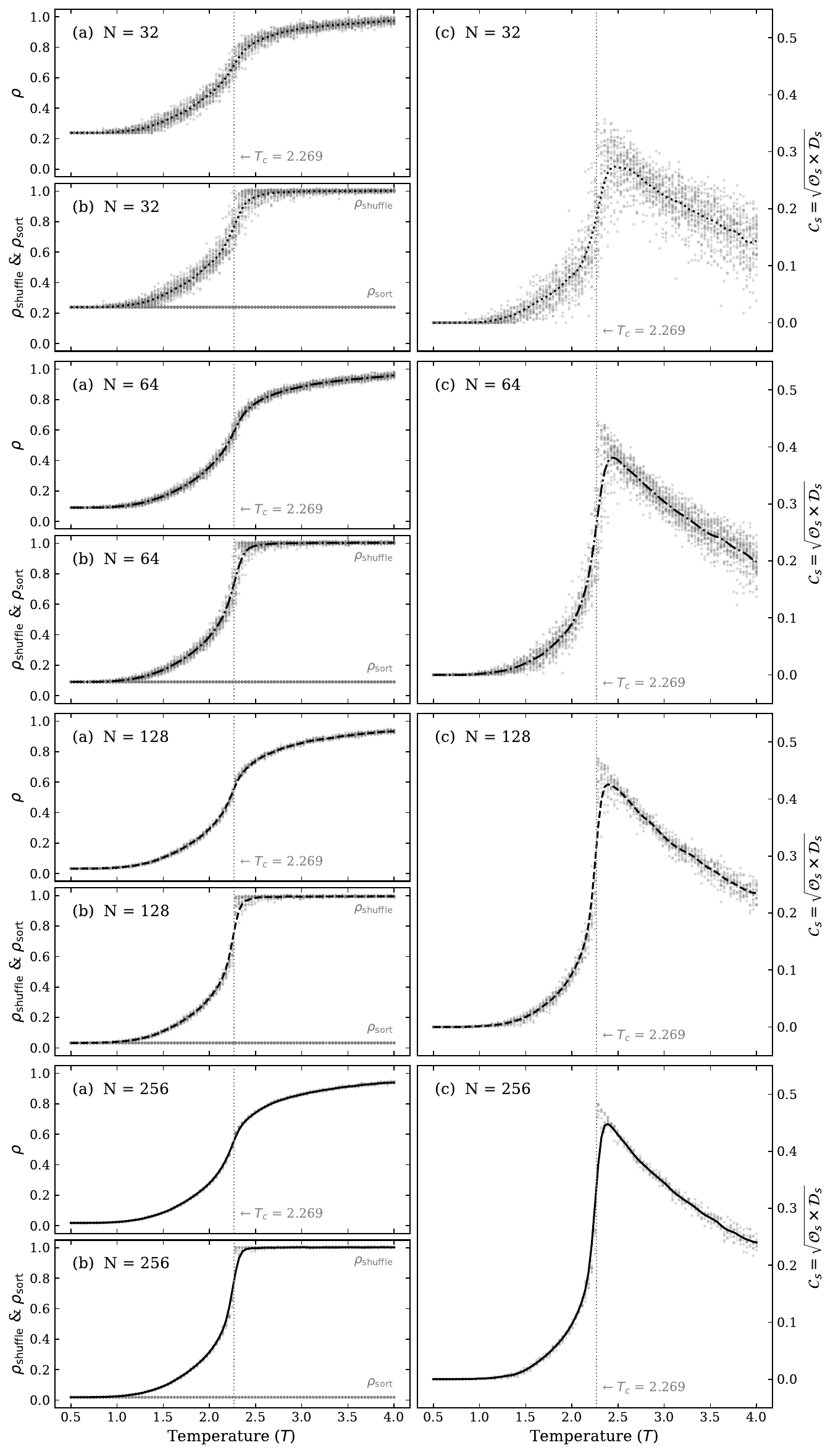}
  \caption{Result of numerical experiments measuring $\mathcal C_s$ using PNG compression for various Ising Lattice sizes, N. \\ Plots show a Gaussian-smoothed mean of the simulated data to highlight the broader trend (Note peak slightly shifted).}
  \label{fig:N_comp_all}
\end{figure*}

\end{document}